\documentclass[%
 aip,
 amsmath,amssymb,
 reprint,%
]{revtex4-1}

\usepackage{graphicx}
\usepackage{dcolumn}
\usepackage{bm}

\usepackage[utf8]{inputenc}
\usepackage[T1]{fontenc}
\usepackage{mathptmx}
\usepackage{etoolbox}

\usepackage{xcolor}
\usepackage{amsmath}
\usepackage{amsfonts}
\usepackage{amssymb}
\usepackage{siunitx}

\usepackage{hyperref}
\hypersetup{
    colorlinks=true,
    linkcolor=blue,
    filecolor=blue,      
    urlcolor=blue,
    citecolor=blue,
    }

\usepackage[normalem]{ulem}
\usepackage{tikz}
\usetikzlibrary{shapes.geometric, arrows}

\newcommand{\REV}{\textcolor{black}}

\makeatletter
\def\@email#1#2{%
 \endgroup
 \patchcmd{\titleblock@produce}
  {\frontmatter@RRAPformat}
  {\frontmatter@RRAPformat{\produce@RRAP{*#1\href{mailto:#2}{#2}}}\frontmatter@RRAPformat}
  {}{}
}%
\makeatother

\begin{document}


\title{Contact mechanics and friction of soft materials: An instrument combining multi-axes dynamical actuation/measurement and \textit{in situ}/\textit{in operando} visualization} 



\author{Matthieu Guibert}
\author{Antoine Aymard}
\author{Cristobal Oliver}
\author{Thibaut Durand}
\author{Baptiste Boulet}
\author{Djibril Gabriel Kashala}
\author{Antoine Mille}
\affiliation{CNRS, Ecole Centrale de Lyon, ENTPE, LTDS, UMR5513, 69130 Ecully, France}
\author{Mathieu Leocmach}
\author{Loïc Vanel}
\affiliation{Université Lyon 1, CNRS, ILM, UMR 5306, F-69100 Villeurbanne, France}
\author{Davy Dalmas}
\author{Julien Scheibert}
\email[Correspondance should be addressed to: ]{matthieu.guibert@ec-lyon.fr ; julien.scheibert@cnrs.fr}
\affiliation{CNRS, Ecole Centrale de Lyon, ENTPE, LTDS, UMR5513, 69130 Ecully, France}



\date{\today}

\begin{abstract}
The mechanics and friction of contact interfaces involving soft materials like gel, rubber or human skin are of both fundamental and applied interest. Recent insights have been made into this field thanks to \textit{in situ} observations of the contact interface. However, current soft-material--oriented tribometers enable only few degrees of freedom for the actuation of the contact, far from covering the richness of the loading conditions relevant to real tribological contacts. Here, we introduce an instrument dedicated to the study of the contact mechanics and friction of soft materials which, in addition to \textit{in situ / in operando} optical monitoring of the interface, enables simultaneous actuation along five degrees of freedom: three translations and two rotations. While all three translations feature large velocity/large stroke motion, the one responsible for normal contact loading can also apply high frequency/small amplitude vibrations. The contact's dynamical response is monitored using both a 6-axes force/torque sensor and a 6-axes displacement/rotation sensor. We first describe the structure of the instrument, its implementation, alignment, calibration and resolutions. We then illustrate its capabilities through a series of experiments on elastomer contacts. Our instrument will be useful to investigate the mechanics of a wide range of soft interfaces submitted to rich, tribology-relevant kinematic or dynamic stimuli.\\

\noindent Now published as: M. Guibert, A. Aymard, C. Oliver, T. Durand, B. Boulet, D. Gabriel Kashala, A. Mille, M. Leocmach, L. Vanel, D. Dalmas, J. Scheibert; Contact mechanics and friction of soft materials: An instrument combining multi-axes dynamical actuation/measurement and in situ/in operando visualization. \textit{Rev. Sci. Instrum.} \textbf{97}, 103905 (2026). \url{https://doi.org/10.1063/5.0351616}
\end{abstract}


\maketitle 

\section{Introduction}

Contact interfaces involving soft materials are ubiquitous. For instance, the elastomers of shoe soles~\cite{worobets_influence_2015}, tires~\cite{tuononen_digital_2014}, artificial fingers~\cite{scheibert_role_2009} or robotic grasping tools~\cite{spiers_variablefriction_2018} are used everyday for their high friction stress against a variety of counter-surfaces. Also, contact with our tongue and skin is central to the tactile perception of food inside the mouth~\cite{dewijk_role_2003} and of surface roughness~\cite{sahli_tactile_2020}, respectively. Yet, the mechanical and frictional properties of such contacts are complex and remain insufficiently understood (see e.g. Refs.~\cite{vakis_modeling_2018, weber_experimental_2022, scheibert_measuring_2027} for recent reviews). The scientific challenges stem from an intricate combination of various distinctive behaviours of both the bulk and interface of soft materials.

First, due to their high compliance, soft materials typically establish large contact areas~\cite{muser_meeting_2017}, which significantly contribute to the substantial friction forces observed~\cite{sahli_evolution_2018}. Moreover, these large contact areas make them particularly amenable to \textit{in situ / in operando} imaging, fostering numerous advancements in the field through direct observation of the interface (see e.g. Refs.~\cite{baumberger_selfhealing_2003,nase_pattern_2008,waters_modemixitypddependent_2010,prevost_probing_2013,delhaye_dynamics_2014,dalbe_multiscale_2015,mcghee_contact_2017,sahli_evolution_2018,sahli_shearinduced_2019, scheibert_onset_2020,xu_asperitybased_2022}).

Second, soft materials are often viscoelastic, which is responsible for, among others things, velocity-dependent friction~\cite{putignano_experimental_2013,scaraggi_friction_2015} and aging~\cite{shoaib_influence_2019}. To better understand the role of viscosity on a soft contact, it is useful to vibrate it with controlled frequencies, amplitudes, waveforms and/or directions (compression and/or shear). Experiments in which soft contacts are vibrated have investigated either the viscoelastic moduli of the contacting materials~\cite{yin_dynamic_2004,boyer_dynamic_2009,samadi-dooki_indirect_2017}, or the effects of viscoelasticity on the contact properties, e.g. its stiffness~\cite{wahl_oscillating_2006}, \REV{adhesion~\cite{charrault_experimental_2009,tricarico_enhancement_2025, yi_pdms_2025, tricarico_influence_2026} or friction~\cite{chen_ultrafast_2026}}. In the above mentioned studies, vibrations were applied along a single direction, normal to the contact, \REV{whereas vibrations along the tangential directions can provide information on the lateral stiffness~\cite{carpick_lateral_1997}, or be used to  actively control friction~\cite{benad_active_2019}.}

Third, the current state of a soft contact is strongly loading-path--dependent, due to adhesion~\cite{dorogin_role_2017} and/or friction~\cite{aleshin_solution_2016}. To investigate such memory effects, it is thus desirable to be able to apply complex, multi-axes stimuli to the contact. Such capability would for instance enable experimental tests of theoretical predictions~\cite{popov_relaxation_2015,aleshin_solution_2016} or reproduction of realistic loading histories relevant to various applications, including tire wear~\cite{mane_new_2013,chanal_characterization_2025, huang_new_2018}, gecko locomotion~\cite{gravish_frictional_2008} or haptics~\cite{ha_full_2025}.
The available loading histories may, for example, enable investigating the role of significant negative normal loads, which can occur due to the combination of a large adhesion and a small elastic modulus, and under which friction forces continue to be active~\cite{mergel_continuum_2019}.

In this context, the scope of this article is to introduce and validate a new contact-mechanics--oriented instrument capable of addressing a broader range of soft-contact--related scientific challenges than any single instrument currently available in the literature. As we will demonstrate, our remotely operable and programmable instrument features:
(i) \REV{actuation} of a soft contact along five independent degrees of freedom simultaneously, \REV{including vibrations along the normal to the contact}, with three control modes \REV{(open loop, force closed loop, position closed loop)},
(ii) real-time monitoring of the six displacements/rotations of one of the two solids in contact,
(iii) measurement of the six components of the forces/torques applied at the contact, and
(iv) \textit{in situ / in operando} imaging of the contact interface.

In this paper, we will describe the design and technical choices used to build the instrument (section~\ref{description}), its alignment and calibration (section~\ref{sec:calibration}), before illustrating its capabilities through a series of experiments involving various elastomer-based contacts (section~\ref{sec:illustration}). 

\section{Description of the instrument}\label{description}

\subsection{Structure}\label{structure}

\begin{figure*}[htb!]
    \centering
    \includegraphics[width=1.99\columnwidth]{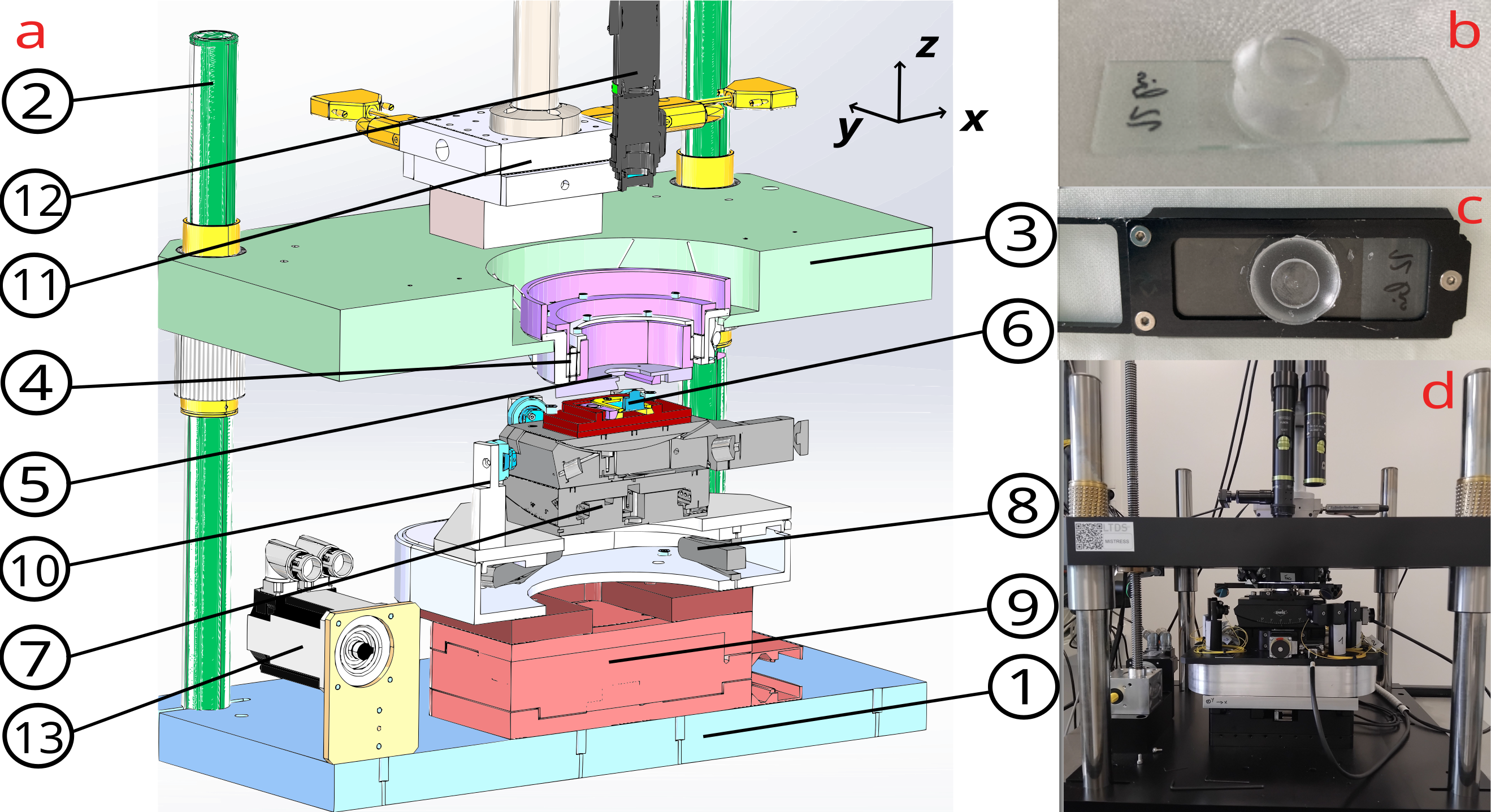}
    \caption{\REV{(a):} Sketch of the instrument. (1) Supporting table, (2) guiding rods, (3) top plate, (4) 6-axes force/torque sensor, (5) upper sample holder, (6) lower sample holder, (7) $\theta_\text{x}\theta_\text{y}$ goniometers, (8) z piezoelectric translation stage, (9) xy linear brushless servomotor translation stage, (10) 6-axes interferometer-based position sensor, (11) xy DC servomotor translation stage for the (12) camera and (13) motor enabling vertical motion of the top plate. \REV{(b): Single spherical elastomer sample with radius of 9.42\,\unit{\milli\meter} (used in Secs.~\ref{sec:Friction} and \ref{sec:vibration}), molded onto a standard microscope glass plate. (c): Sample mounted in the sample holder. (d): Photograph of the instrument.}}
    \label{fig:schema_mistress}
\end{figure*}

Figure~\ref{fig:schema_mistress} illustrates the mechanical layout/design of the instrument, with the panel d showing a photograph of the actual device. The instrument is designed around a pair of sample holders (labels 5 and 6 in Fig.~\ref{fig:schema_mistress}), in which \REV{glass slides of size 76\unit{\milli\meter}$\times$26\unit{\milli\meter}$\times$1.2\unit{\milli\meter} and bearing the} tribological surfaces of interest can be fixed.

The lower sample holder (label 6) is placed on top of a stack of motorized actuators. Starting from the supporting table (1), a high precision xy linear servomotor translation stage (9 ; Allio AI-LM-10000-XY) enables translations along the $\mathbf{x}$ and $\mathbf{y}$ axes (see axes definition in Fig.~\ref{fig:schema_mistress}, top) in the range $\pm 50$\,\unit{\milli\meter} with an accuracy $\pm 3$\,\unit{\micro\meter}, a repeatability of $\pm 30$\,\unit{\nano\meter}, and a speed ranging from 1\,\unit{\micro\meter\per\second} up to 0.8\,\unit{\meter\per\second}. On top of this stage, a laboratory-made high-load high-stroke capacity piezoelectric translation stage (8) is mounted. It consist of  three piezoelectric actuators (Cedrat APA 1000L, \REV{maximum push force of 350\,\unit{\newton} each}) placed at 120\,\unit{\degree} on a circle of radius 105\,\unit{\milli\meter}, and fixed between two rigid plates. The stage is controlled by a single amplifier (Piezosystemjena ENT40) and enables \REV{vibration and} high-precision motion (repeatability better than 10\,\unit{\nano\meter}) of amplitude up to 900\,\unit{\micro\meter} along the $\mathbf{z}$ axis. Between this piezoelectric table (8) and the lower sample holder (6), a double stepper-motorized goniometer with both rotations sharing the same center (7; OWIS, TPM 150-20-20-243) is placed and enables rotations of the lower sample around the $\mathbf{x}$ and $\mathbf{y}$ axes in the range $\pm 10$\,\unit{\degree}, with a repeatability better than 0.01\,\unit{\degree} per axis.
\REV{In this configuration, i.e., when loaded by the goniometer (7; 5.9\,\unit{\kilo\gram}) and the sample holder (6; 0.5\,\unit{\kilo\gram}), the piezoelectric table (8) has a remaining push force of about 950\,\unit{\newton}, and its first resonance frequency is at 43$\pm$2\,\unit{\hertz}. In the tested frequency range of $f$=[2-25]\,\unit{\hertz}, the maximum peak-to-peak amplitude of a sinusoidal $z$-displacement oscillation that can be produced by the piezoelectric table is $b f^{-k}$, where $k$=0.79$\pm$0.09 and $b$=353$\pm$66\,\unit{\micro\meter.\hertz^{k}}.}

The upper sample holder (5) is attached to the lower end of a custom 6-axes force/torque sensor (4), fully described in Ref.~\cite{guibert_versatile_2021}, which measures the three orthogonal forces (along $\mathbf{x}$, $\mathbf{y}$ and $\mathbf{z}$) with resolutions of 0.3, 0.4, and 0.1\,\unit{\milli\newton}, respectively, and the three torques around the same axes with resolutions of 1.0, 1.2, and 1.2\,\unit{\micro\newton}.\unit{\meter}, respectively. The sensor stiffnesses are \REV{174$\pm$22, 215$\pm$6 and 144$\pm$4}~\unit{\kilo\newton}/\unit{\meter} along $\mathbf{x}$, $\mathbf{y}$ and $\mathbf{z}$, \REV{483$\pm$25 and 409$\pm$15}~\unit{\newton}.\unit{\meter}/\unit{\radian} for rotations around $\mathbf{x}$ and $\mathbf{y}$, and \REV{1588$\pm$26}~\unit{\newton}.\unit{\meter}/\unit{\radian} for rotations around $\mathbf{z}$. \REV{The smallest of these measured stiffnesses, together with the mass of 1.035\,\unit{\kilo\gram} carried by the force sensor (lower part of the sensor and upper sample holder) translate into a resonance frequency of about 59\,\unit{\hertz}. This value is consistent with the result of finite element calculations (SOLIDWORKS Simulations 2024 SP5.0, modal analysis with direct solver, 621356 tetrahedral elements of size down to 100\,\unit{\micro\meter}), which give a first resonance frequency of 84.5\,\unit{\hertz}, corresponding to a vibration along the $z$-axis. The maximum measurable force in the vertical direction is currently about 10\,\unit{\newton}, a value that may be adjusted by modifying either one of two inner components of the force/torque sensor: the measurement range of its capacitive displacement sensors or the stiffness of the elastic blades that relate its upper and lower ends.}
The sensor is mounted at the bottom of a top plate (3) that can be moved vertically, guided by precision rods and bushings (2 ; Fibro GmbH) mounted on the base plate (1), \REV{using a}  a precision drive (13 ; Kollmorgen AKM31) to coarsely position the samples relative to each other. Note that during experiments, the top plate is kept fixed, so that inertial forces due to accelerations are not affecting the force/torque measurements. The top plate (3) is further equipped with a high-resolution camera (12 ; Teledyne DALSA Genie Nano-GigE, \REV{C-mount USB3 camera with 4112×3008 pixels and maximum frame rate 10\,\unit{\hertz}}) and a motorized zoom lens (Qioptics Optem Fusion) for contact imaging.  This large--working-distance optical system is placed on a xy motorized table (11 ; Newport M-UMR12.40) allowing to center the image on any region of interest. The  optical set up offers images of the contact interface with a pixel size down to 1\,\unit{\micro\meter}/pixel, and a field of view ranging from 3$\times$4 to 9$\times$12\,\unit{\milli\meter\squared}. \REV{Note that the camera could be readily replaced by any other C-Mount camera with different resolution, color or frame rate.}
This visualization device enables to monitor \textit{in situ / in operando} the contact interface from above thanks to the hollow structure of the top part of the instrument (top plate, force/torque sensor and upper sample holder).

The six-degrees-of-freedom relative motion (three translations along and three rotations around the $\mathbf{x}$, $\mathbf{y}$, and $\mathbf{z}$ axes) between the upper sample holder and the top plane of the piezoelectric table is measured using six interferometers (10 ; Attocube IDS-3010) arranged in a Stewart-platform configuration~\cite{stewart_platform_1965}. The interferometers are positioned in pairs at the ends of three posts mounted on the plane connecting the piezoelectric stage (8) and the goniometers (7), at the vertices of an equilateral triangle, and are oriented at 45\unit{\degree} with respect to the $\mathbf{z}$ axis. The beam of each interferometer points on a precision flat mirror with planeity better than $\lambda /20$, fixed on the upper sample holder (5). This sensor enables measurements of translation amplitudes in the range $\pm 10$\,\unit{\milli\meter} with a resolution better than 1\,\unit{\nano\meter}, and of rotation angles in the range $\pm$ \ang{0.2} with a resolution better than \ang{0.0002}. Note that the measured motion is unaffected by rotations imposed by the goniometers.

\subsection{FPGA-based processing}

\begin{figure*}[ht!]
    \centering
    \includegraphics[width=1.99\columnwidth]{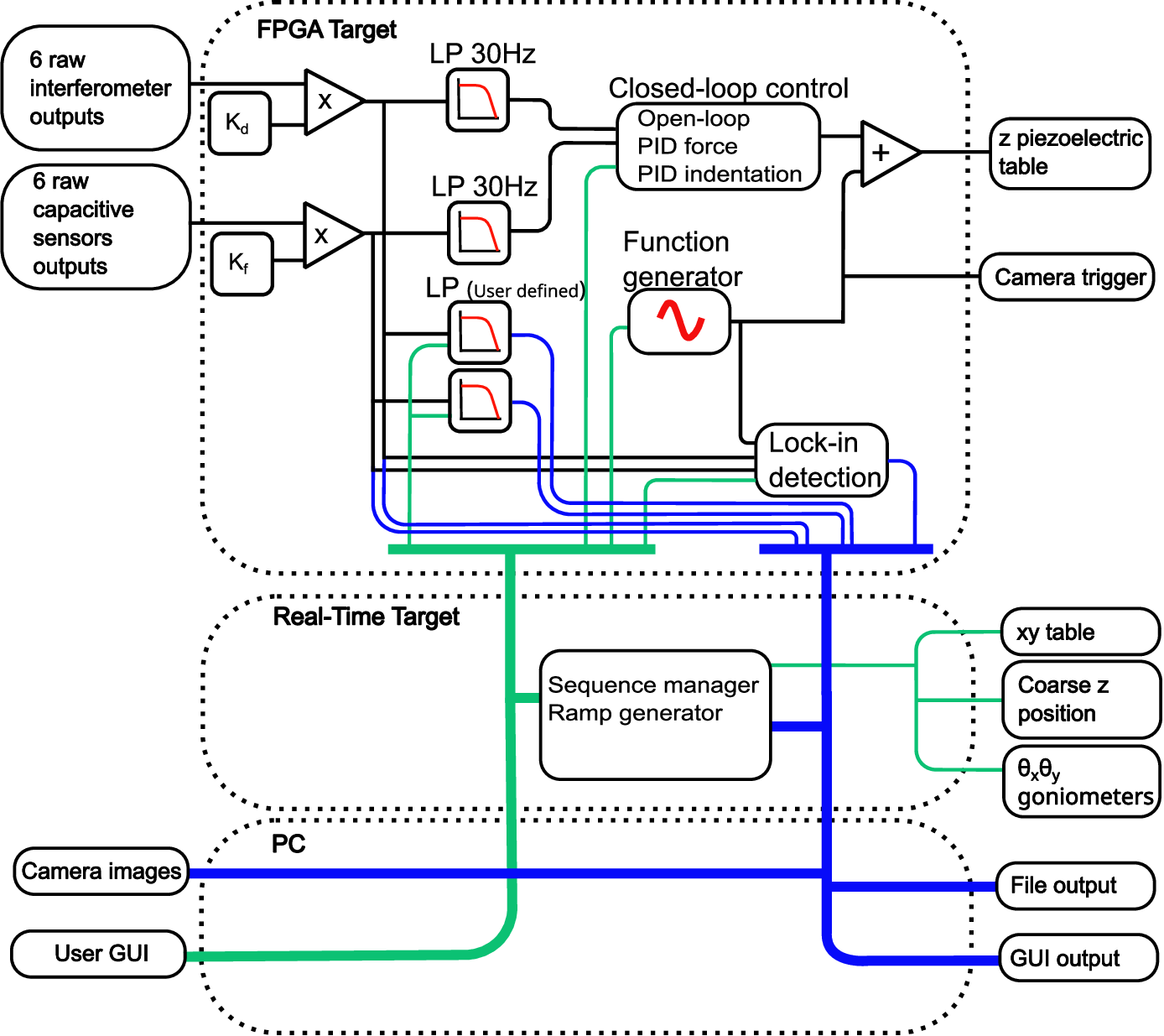}
    \caption{
    Schematic of the FPGA, real-time, and PC control levels of the instrument. PC level (bottom): user interface for entering action sequences and visualizing contact images and force and displacement components. Real-time target (middle): Inputs and outputs are processed at 100\,\unit{\hertz} to generate control signals for all actuators except the piezoelectric table. It also integrates the sequence manager and a ramp generator. FPGA target (top): Force and displacement components are computed from raw sensor outputs (using matrices $K_\text{f}$ and $K_\text{d}$) and expressed in the ($\mathbf{x}$, $\mathbf{y}$, $\mathbf{z}$) frame at 3\,\unit{\kilo\hertz}. These components are low-pass filtered at 30\,\unit{\hertz} (upper filters) before entering the feedback loop, balancing precision and stability, and are also filtered at a user-defined frequency (lower filters) for data logging and display. The feedback loop output drives the piezoelectric table. An embedded lock-in amplifier applies a sinusoidal motion to the piezoelectric table at a defined frequency and performs lock-in detection of the force/displacement amplitude and phase components at that frequency.}
    \label{fig:shema_control}
\end{figure*}

The overall architecture of the real-time software is described in Fig.~\ref{fig:shema_control}. The raw outputs from the force sensor (six capacitive displacement sensors MCC10, conditioned with Fogale MC900) are acquired via six 24-bit delta-sigma analog-to-digital converters (NI9239). The 1$\times$6 vector signal is then multiplied by a 6$\times$6 force matrix $K_\text{f}$ (see section~\ref{sec:calibforce}) to give the 1$\times$6 force vector (three forces and three torques). Simultaneously, the raw outputs from the displacement sensor (the six interferometers mentioned in section~\ref{structure}) are in A-quad-B form and are acquired through two NI9401 acquisition counters. This 1$\times$6 vector is multiplied by a 6$\times$6 displacement matrix $K_\text{d}$ (see section~\ref{sec:calibdisp}) to give the relevant 1$\times$6 displacement vector (three displacements and three rotation angles).

To perform the above-mentioned matrix multiplications at a sufficiently high rate for  closed-loop control of the piezoelectric stage, a field-programmable gate array (FPGA, NI-CRIO 9037) is used. It allows not only these critical calculations, but also filtering or lock-in amplification (see in the following) to be executed at 3\,\unit{\kilo\hertz} with deterministic timing. 

Force and displacement signals are first low-pass (LP) filtered at a user-defined cutoff frequency and sent to the personal computer (PC, see the bottom part of Fig.~\ref{fig:shema_control}) for recording (file output). In parallel, they are also LP-filtered using a cascaded integrator comb filter ensuring less than 0.1~\unit{\decibel} loss between 0 and 30~\unit{\hertz}, and 120~\unit{\decibel} reduction after 60~\unit{\hertz} (see LP 30~ \unit{\hertz} blocks in Fig.~\ref{fig:shema_control}) before being fed into the closed-loop control function. This function generates the drive signal for the vertical position of the piezoelectric actuator via a 24-bit digital-to-analog converter (NI-9260). Closed-loop control can operate in three modes:
(i) open-loop mode (no feedback),
(ii) closed-loop constant normal force mode (the piezoelectric stage adjusts in real time to maintain a prescribed normal force), or
(iii) closed-loop constant normal \REV{displacement} mode (the piezoelectric stage adjusts in real time to maintain a fixed separation between the sample holders).
Each mode employs a dedicated set of proportional-integral-derivative (PID) controller parameters.

Before reaching the digital-to-analog converter that drives the piezoelectric stage, a sinusoidal numerical signal created by a function generator may be added to the PID loop output. This additive signal, whose angular frequency, $\omega$, and amplitude, $A$, are user-controlled, can be used to impose a known dynamic excitation on the tribological contact. In this case, the contact's response to such excitation is extracted from the raw force and/or displacement signals using a two-phase lock-in amplifier~\cite{meade_advances_1982,zhang_lockin_2024}. This function multiplies a measured signal, $S(t)$, by reference sine and cosine waves, $A\sin(\omega t)$ and $A\cos(\omega t)=A\sin(\omega t+\pi/2)$, and low-pass filters both results using a cascaded integrator comb filter. \REV{This filter is fully user-defined, with an adjustable cutoff frequency ranging from 0.1 to 5~\unit{\hertz}, and an adjustable attenuation of 60, 90 or 120\,dB beyond the cutoff frequency. These values correspond to a response time of the filter ranging from 100~\unit{\milli\second} to 2.5~\unit{\second}.}  The overall process yields two scalars, $X_\omega$ and $Y_\omega$, respectively. The amplitude, $S_\omega=\sqrt{X_\omega^2+Y_\omega^2}$, of the sine component of $S$ at angular frequency $\omega$, and the phase shift of this component with respect to the reference sine wave, $\theta=tan^{-1}(Y_\omega/X_\omega)$, are then calculated. These calculations are performed for both force and displacement signals. Note that implementing them on the FPGA target~\cite{meade_advances_1982,galaviz-aguilar_fieldprogrammable_2025} preserves the full resolution of the raw signals: 24 bits for force signals (10\,V full scale) and 32 bits with a resolution of 0.1\,\unit{\nano\meter} for displacement signals.

\subsection{Real-time control for motions and measurements}

Once two surfaces of interest are inserted into the two sample holders, all steps of a contact mechanics measurement campaign can be performed and controlled via a custom LabVIEW interface on a standard PC. This interface enables direct interaction with all motorized components of the device (mechanical and optical), as well as with all control parameters (including calibration matrices, PID parameters, filter settings and lock-in parameters), facilitating easy tuning of all functionalities. All parameters are managed through human-readable files and stored under version control, ensuring full traceability of all experiments conducted over the instrument's lifetime. In addition to handling input parameters, the software manages the acquisition and storage of displacement, force and imaging data, while providing live visualization of these outputs during operations.

Communication between the LabVIEW program and the hardware is performed via a real-time controller that supports two operational modes.
First, in \textit{manual mode}, each motorized element can be controlled individually by applying direct instructions. This mode is particularly useful for preparing measurement campaigns, allowing users to, e.g., perform a coarse approach of the two sample holders near initial contact, align the samples, and adjust camera settings (including field of view, focus and exposure time).
Second, in \textit{automated mode}, users can program and execute arbitrarily long sequences of pre-defined actions including motions, image acquisitions, force or displacement ramp generation, all with fine control over timing jitter (see the real-time target in Fig.~\ref{fig:shema_control}). The xy translation stage and $\theta_\text{x}\theta_\text{y}$ goniometers (4 axes) are synchronized to enable complex trajectories. The list of possible actions is extensible, allowing users to add custom sequences based on specific needs. In this mode, all motor instructions and closed-loop control along the $\mathbf{z}$ axis operate via a real-time target at 100\,\unit{\hertz}. 

Overall, automated sequences and the possibility to conduct experiments remotely (through a network) both minimize the experiment duration and maximize experimental reproducibility (by avoiding, e.g., vibrations or thermal drift caused by the operator's physical presence). 

\section{Calibration and alignment}\label{sec:calibration}

\subsection{Calibration of the displacements sensor}\label{sec:calibdisp}

The 6$\times$6 matrix $K_\text{d}$, which converts the six output distances of the six interferometers into the three displacements and three rotation angles of the upper sample holder, is obtained from a geometric transformation based on the known positions and orientations of the interferometers, as described in, e.g., Ref.~\cite{hale_principles_1999}. No change of units is involved because the interferometers' outputs are absolute distances to their mirrors. The displacements and angles are expressed in an arbitrary orthonormal frame, $(\mathbf{x}_\text{d},\mathbf{y}_\text{d},\mathbf{z}_\text{d})$, in \unit{\mm} and mrad.

\subsection{Alignment between sensors and actuators}\label{sec:alignment_new}

The physical structure described in section~\ref{structure} involves multiple reference frames: those associated with each displacement system (the xy macroscopic table, the piezoelectric table, and the goniometers), as well as those for each sensor (displacement and force sensors). To facilitate the interpretation of experimental measurements, it is essential to express all measurements in a single, common reference frame.

This alignment involves determining the relative rotations between all frames and correcting them either through mechanical adjustments (when possible) or by applying rotation matrices to the raw measured values. The alignment is performed as follows.

The first step consists in leveling the machine base to ensure it lies in a horizontal plane, i.e., perpendicular to gravity. This is achieved using a digital spirit level (MasterLevel Compact Plus), which allows the base plate to be aligned with an accuracy of 0.05\unit{\degree} relative to gravity. Following this step, the horizontality of the macroscopic motion of the xy translation stage is verified using a flat rule placed on the base and a dial gauge placed on the moving part of the table. Shims are used to ensure that the measured vertical displacements remain below 10\,\unit{\micro\meter} over a $\pm$25\,\unit{\mm} travel range. This allows defining the (orthonormal) actuation frame $(\mathbf{x},\mathbf{y},\mathbf{z})$ with $\mathbf{z}$ being the vertical axis.

The next step is to determine the rotation angles between the actuation frame $(\mathbf{x},\mathbf{y},\mathbf{z})$ and the displacement sensor frame $(\mathbf{x}_\text{d},\mathbf{y}_\text{d},\mathbf{z}_\text{d})$. To achieve this, we first perform four linear motions of 1\,\unit{\milli\meter} amplitude using the xy translation stage along the $\mathbf{x}$, $-\mathbf{x}$, $\mathbf{y}$, and $-\mathbf{y}$ directions, with no contact between the two samples. If the actuation and displacement frames were perfectly aligned, the measured displacements would exhibit negligible components along $\mathbf{y}_\text{d}$ and $\mathbf{z}_\text{d}$ (resp. $\mathbf{x}_\text{d}$ and $\mathbf{z}_\text{d}$) when moving along $\mathbf{x}$ or $-\mathbf{x}$ (resp. $\mathbf{y}$ or $-\mathbf{y}$). However, all components ($x_\text{d}$, $y_\text{d}$, and $z_\text{d}$) are observed to vary linearly, with typical amplitudes of 3\,\unit{\micro\meter} for $z_\text{d}$ over a 1\,\unit{\milli\meter} motion along $\mathbf{x}$. To eliminate these variations, we successively multiply the conversion matrix, $K_\text{d}$, by three rotation matrices:
(i) a rotation around $\mathbf{z}_\text{d}$ by an angle $\psi_\text{d}$, calculated from the measured slope $\frac{\partial x_\text{d}}{\partial y_\text{d}}$,
(ii) a rotation around $\mathbf{x}_\text{d}$ by an angle $\theta_\text{d}$, calculated from the slope $\frac{\partial z_\text{d}}{\partial y_\text{d}}$, and
(iii) a rotation around $\mathbf{y}_\text{d}$ by an angle $\phi_\text{d}$, calculated from the slope $\frac{\partial z_\text{d}}{\partial x_\text{d}}$. After applying these rotations, the displacement components vary by less than 0.05\,\unit{\micro\meter} over 1\,\unit{\milli\meter} displacements, indicating that the actuation and displacement frames are now aligned to within 0.003\unit{\degree}.

Once the displacement sensor frame $(\mathbf{x}_\text{d},\mathbf{y}_\text{d},\mathbf{z}_\text{d})$ is defined, we can assess the verticality of the motion of the piezoelectric translation stage. This is achieved by analyzing the (aligned) displacement measurements over the full travel range of the z piezoelectric stage. The measurements reveal that the parasitic motion of the stage is non-linear across the full scale of motion. To optimize experimental conditions, we shim the base of the piezoelectric stage to achieve the best alignment in the first half of its range, as this portion is the most frequently used in our experiments.

\begin{figure}[htb!]
    \centering
    \includegraphics[width=0.99\columnwidth]{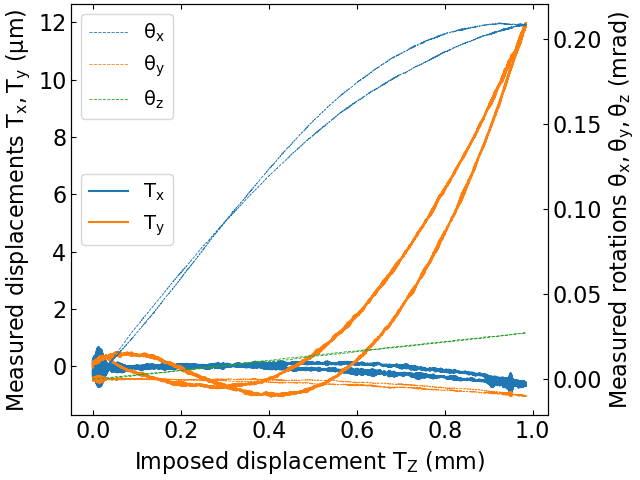}
    \caption{Measured residual displacements ($T_\text{x}$ and $T_\text{y}$, thick solid lines) and rotation angles ($\theta_\text{x}$, $\theta_\text{y}$ and $\theta_\text{z}$, thin dashed lines) during imposed vertical motion of the z piezoelectric translation stage, performed over a back-and-forth stroke over the full possible range, 1\,mm, after alignment. The first half of the motion (the one used in most experiments) maintains parasitic xy deviations of less than 1\,\unit{\micro\meter}, while the maximum deviation over the full stroke remains less than 12\,\unit{\micro\meter}. Rotation angles remain less than about 0.2~\unit{\milli\radian} (0.012\,\unit{\degree}).}
    \label{fig:piezo_alignment}
\end{figure}

The displacement and angle measurements in the aligned conditions, represented in Fig.~\ref{fig:piezo_alignment}, show that the motions along $\mathbf{x}$ and $\mathbf{y}$ remain within 1\,\unit{\micro\meter} for the first 500\,\unit{\micro\meter} of travel and less than 12\,\unit{\micro\meter} over the full 1\,\unit{\milli\meter} stroke. The quasi-linear variation in the angular position of the free end of the piezoelectric stage arises from differences in the linearity of the three individual piezoelectric elements used in the stage and cannot be corrected. These parasitic rotations are measured to be less than about 0.2~\unit{\milli\radian} (0.012~\unit{\degree}).

The final step is to align the force sensor reference frame with the $(\mathbf{x},\mathbf{y},\mathbf{z})$ frame, which is achieved through the sensor calibration process described in the Sec.~\ref{sec:calibforce}.

\subsection{Calibration of the force sensor}\label{sec:calibforce}

\begin{figure}[htb!]
    \centering
    \includegraphics[width=0.99\columnwidth]{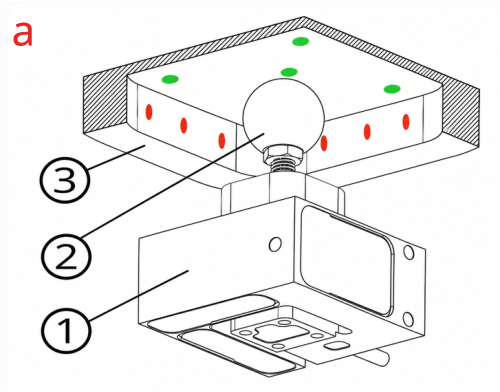}
    \includegraphics[width=0.99\columnwidth]{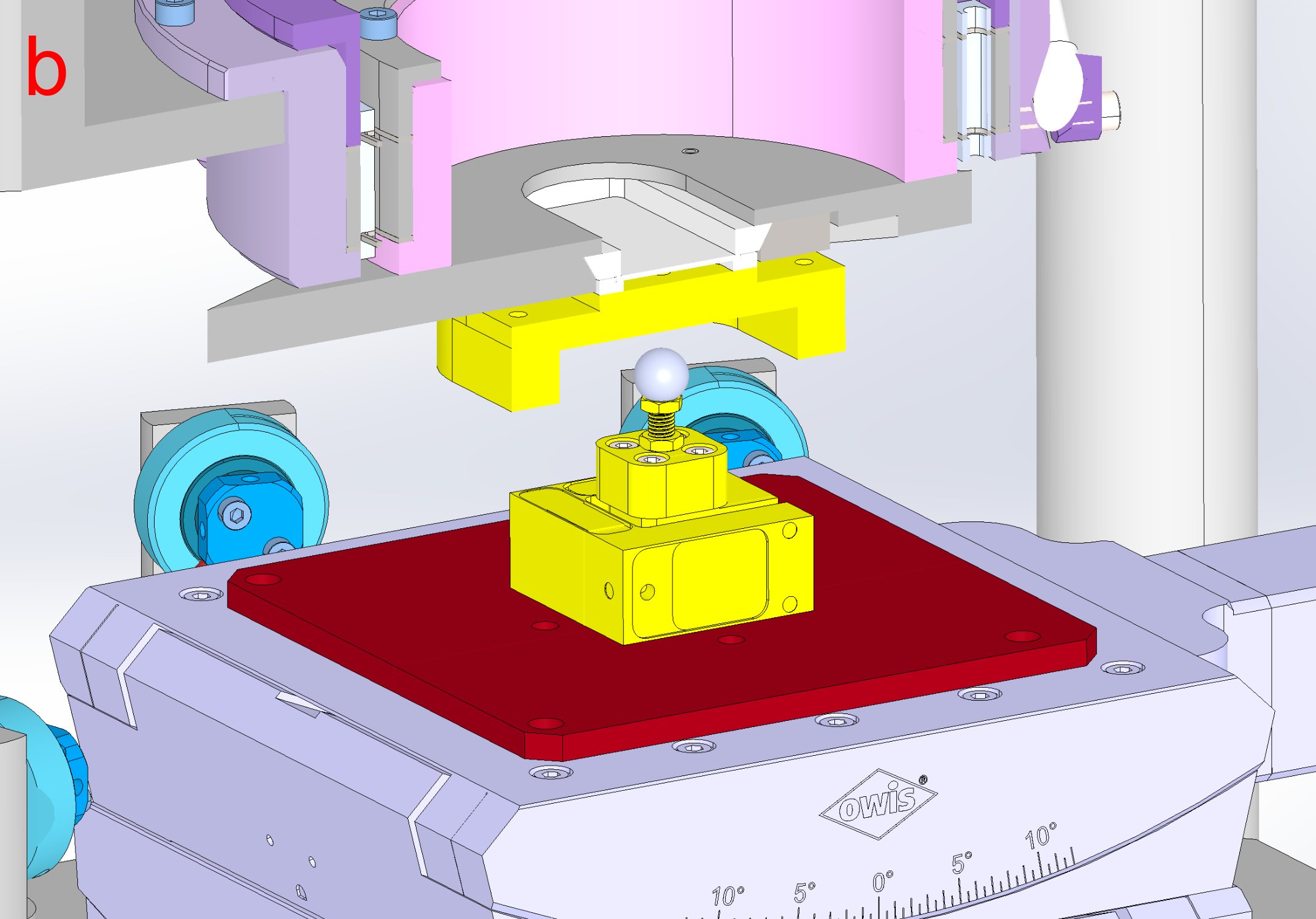}
    \caption{\REV{(a):} 3D view of the force calibration setup. (1) Reference \REV{3-axes} force sensor, (2) steel ball and (3) bell-shaped top part fixed to the 6-axes force sensor to be calibrated. The top part is represented with 2 cut sections, for better reading. The red (resp. green) dots represent the points where a known horizontal (resp. vertical) force is applied. \REV{(b) Detailed view of the calibration set-up (yellow) mounted in the instrument.}}
    \label{fig:Shema_calibration}
\end{figure}

The $6 \times 6$ matrix $K_\text{f}$, which converts the six output voltages of the six capacitive sensors into the three forces and three torques applied to the upper sample holder, is obtained through a calibration procedure described in Ref.~\cite{guibert_versatile_2021} (a typical $K_\text{f}$ matrix for the present sensor is provided in the appendix therein). 
For the specific calibration performed here, known forces are measured simultaneously by both a 3-axes reference force sensor (K3D40, PM Instrumentation) and the current 6-axes sensor. The forces are applied between a steel ball placed on the moving part of the instrument and a bell-shaped aluminum part placed at the free end of the 6-axes force sensor to be calibrated (see Fig.~\ref{fig:Shema_calibration}). The reference sensor is positioned under the steel ball and aligned with the actuation frame $(\mathbf{x}, \mathbf{y}, \mathbf{z})$ using the goniometers. This alignment ensures that a dead weight applied to the reference sensor produces a force only along the $\mathbf{z}$ axis, with no projected forces along the $\mathbf{x}$ and $\mathbf{y}$ axes.

Using the xy translation stage, the steel ball is brought into contact with the bell-shaped part at different locations (three on each of the four lateral sides, see the red points in Fig.~\ref{fig:Shema_calibration}, and 5 on the top face, see green points), enabling application of forces along the $\mathbf{x}$, $\mathbf{-x}$, $\mathbf{y}$, $\mathbf{-y}$, and $\mathbf{z}$ directions with varying torques. For each contact point, the applied force is measured by the reference force sensor, while the torques are calculated using the position of the xy stage. At each point, 12 different force amplitudes ranging from 0.1 to 5\,\unit{\newton} are applied. Full unloading is performed between each measurement to avoid hysteretic effects. The linearity and absence of hysteresis are verified independently for each set of points used in the full calibration.

We emphasize that any force component tangential to the contact (due, e.g., to friction or angular misalignment of the bell-shaped part) are captured by the 3-axes reference sensor and thus fully accounted for in the calibration. The computed matrix $K_\text{f}$ is the one that best captures, in the least-squares sense, the relationship between all pairs of $1\times6$ force/torque vectors measured thanks to the reference sensor and $1\times6$ output voltage vector from the 6-axes force sensor to be calibrated. By construction, $K_\text{f}$ is expressed in an orthonormal frame identical to the actuation frame $(\mathbf{x}, \mathbf{y}, \mathbf{z})$.

At the end of the calibration we  verified, for each of the 204 measurements (12 force amplitudes $\times$ 17 contact locations), that the difference between the forces and torques given by the reference sensor and by the calibrated 6-axes sensor always remains less than 3\,\unit{\milli\newton} and 0.15\,\unit{\milli\newton}.\unit{\meter}, respectively.

\section{Illustration examples}\label{sec:illustration}
In this section, we present three experiments that illustrate compelling capabilities of our instrument to address key requirements commonly encountered in contact mechanics and tribological studies. First, we show how to leverage synchronized multi-axes motion and contact visualization to precisely align two surfaces. Second, we showcase the ability to perform well-controlled friction experiments (under either constant force or constant displacement) thanks to FPGA-implemented closed-loop control. Third, we leverage piezoelectric-enabled vertical vibrations to perform accurate detection of the first contact between surfaces. \REV{For all experiments, the instrument is installed in a clean room with monitored environment (temperature 22$\pm${2}\unit{\degree}C and humidity 50$\pm${25}\%)}.

\subsection{Multi-axes motion and sample alignment capability}\label{sec:imaging}

\begin{figure*}[ht!]
    \centering
    \includegraphics[width=1.99\columnwidth]{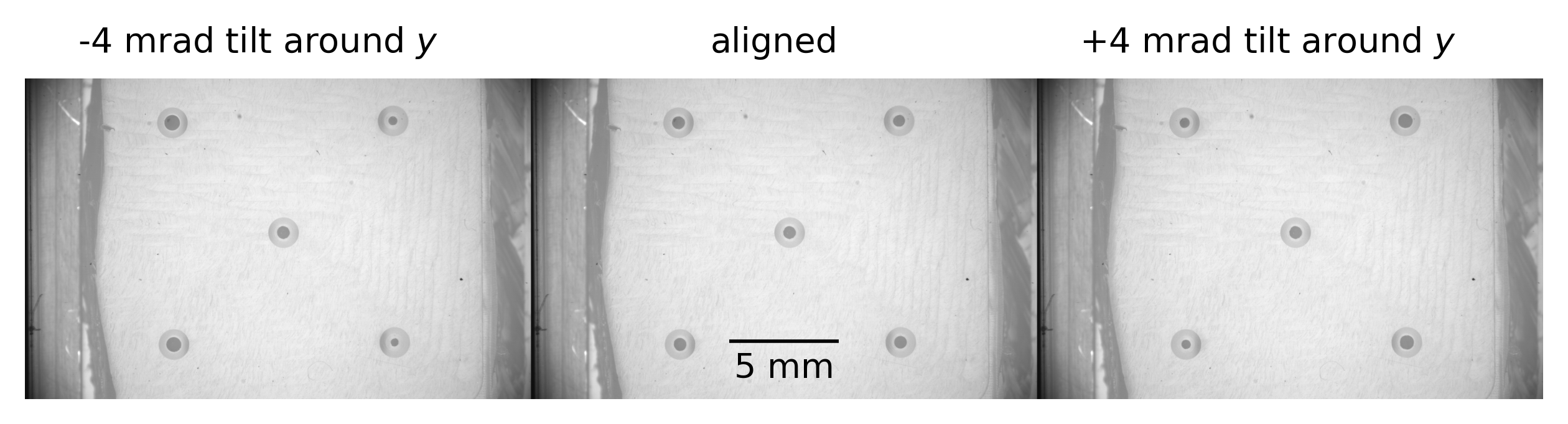}
    \caption{Images of the contact between a smooth glass plate and a nominally flat PDMS sample decorated with five spherical asperities (radius of curvature $1502 \pm 21\,\unit{\micro\meter}$ and summit height with respect to the sample's base plane $173\,\unit{\micro\meter}$), under a normal force 0.5\,\unit{\newton}. The four corner asperities form a square with a size of 10\,\unit{\milli\meter}. From left to right: the PDMS sample is tilted by -4, 0 and 4\,\unit{\milli\radian} with respect to the aligned configuration. The real contact area on each microcontact corresponds to the inner dark circle, and depends on the tilt angle.}
    \label{fig:3_angles_barycenter}
\end{figure*}

Alignment of samples is critical in many tribological experiments, particularly for flat-on-flat contact interfaces, including those relevant to frictional metainterfaces~\cite{aymard_designing_2024,zeka_normal_2026,fu_automated_2026,scheibert_slope_2027}. These metainterfaces are calibrated using planar samples decorated with five identical spherical asperities, and accurate preliminary alignment is required to ensure that all five asperities create identical contacts with a rigid smooth indenter. Alignment is validated thanks to contact visualization by checking that the contact areas of all five contacts are equal.

Such an alignment process is tedious as soon as the goniometers rotate around a center that is not exactly located on the contact interface, which is generally the case. Indeed, if a vertical distance $d_\text{z}$ exists between the rotation center and the tribological surface, any small rotation $\theta_\text{y}$ around the $\mathbf{y}$ axis will induce an $\mathbf{x}$-displacement of the sample, $d_X=\theta_\text{y} d_\text{z}$. Such displacement will shear the contact and affect the contact areas~\cite{papangelo_shearinduced_2019,lengiewicz_finite_2020,zhang_nonmonotonic_2024}, preventing any assessment of the alignment quality, unless one breaks and recreates the contact. Here, to facilitate the alignment procedure, we leverage the possibility of our instrument to impose custom multi-axes motions: simultaneously to a rotational motion $\theta_\text{y}(t)$, we apply a translational motion along $\mathbf{x}$ of amplitude $-\theta_\text{y}(t) d_\text{z}$ that compensates the rotation-induced unwanted translation. Here, $d_\text{z}$ is first calibrated by measuring the $\mathbf{x}$-displacement induced by a reference rotation angle, while the sample is shifted vertically by a know distance sufficient to bring the two surfaces out-of-contact.

To test this strategy, we use a smooth and flat glass plate (upper sample holder) as well as a nominally flat silicon elastomer sample (polydimethylsiloxane, PDMS, Sylgard 184, cross-linked as recommended in Ref.~\cite{delplanque_solving_2022}, lower sample holder) decorated with five spherical asperities of identical radius of curvature ($1502 \pm 21\,\unit{\micro\meter}$) and summit heights with respect to the sample's base plane ($173\,\unit{\micro\meter}$).
Here, we focus solely on alignment involving rotation about the $\mathbf{y}$ axis, i.e., using only one of the two goniometers. However, the same procedure is applicable (and mandatory to achieve perfect alignment) along the $\mathbf{x}$ axis using the second goniometer. $d_\text{z}$ was evaluated to be 132.821~\unit{\milli\meter}, based on a lateral displacement of 1.33~\unit{\milli\meter} (measured on the interface images) caused by a rotation of 10~\unit{\milli\radian}. After applying the $-\theta_\text{y}(t) d_\text{z}$ correction, the remaining parasitic displacement of the sample, observable from the interface images, is lower than 2~\unit{\micro\meter}, demonstrating that the total generated motion (translation along $\mathbf{x}$ plus $\theta_\text{y}$ goniometer rotation) is a pure rotation of the sample with a center located on its surface.

This is illustrated in Fig.~\ref{fig:3_angles_barycenter}, which shows three contact configurations with the same normal force, but three different angles of the PDMS sample with respect to the glass plate. The in-plane difference in the locations of the contact centers between these images is smaller than $1$\,\unit{\micro\meter}. The middle image corresponds to the aligned case, where the five contact areas are almost equal ($0.250\pm0.013$\,\unit{\milli\meter\squared}).  The side images correspond to tilted configurations where the misalignment around the $\mathbf{y}$-axis is visible through the difference between the areas of the left and right contacts, while the area of the central contact is unchanged.

\subsection{Force and displacement closed-loop control}\label{sec:Friction}

In many mechanical tests, and particularly in tribology experiments, there are two primary control modes: displacement-controlled or force-controlled. \REV{A given contact can behave differently according to which mode is used to actuate it. A classical example is the measurement of the pull-off force (the maximum tensile force necessary to detach two surfaces in contact) of an elastic adhesive contact between a sphere and a plane~\cite{maugis_adhesion_1992}. For short-ranged adhesive interactions, the pull-off force under force control is about twice that under displacement control. For long-ranged interactions, the difference is even qualitative: detachment occurs through a jump instability under force control while the contact area vanishes in a stable way under displacement control (see Fig.~8 in Ref. \cite{maugis_adhesion_1992}). Thus the capability to actuate a contact in either of the two modes can give access to non-trivial information about the interface (in the above-mentioned example, the range of adhesive interactions).}

To demonstrate the capability of our instrument to perform tests under both types of conditions, thanks to FPGA-implemented closed-loop control, we conducted a series of tribological tests between a PDMS hemisphere (Sylgard 184, radius of curvature 9.42\,mm, placed in the upper sample holder) and a glass plate (lower sample holder). We first aligned the glass surface with the $(\mathbf{x}, \mathbf{y})$ plane by (i) measuring, at various $(T_\text{x}, T_\text{y})$-locations of the contact along the glass, the vertical position $T_\text{z}$ of the piezoelectric stage necessary to reach a common normal force of 0.1\,N with the sphere and (ii) tilting the two goniometers by the angles necessary to compensate the measured inclination of the plane formed by all measured $(T_\text{x},T_\text{y},T_\text{z})$ points. Note that this sample alignment procedure is different from the multi-contact-based one used in section~\ref{sec:imaging} because here we have only one contact. Then, we intentionally tilted the glass plate by known angles (from $-5$ to $5\,\unit{\milli\radian}$ in steps of $2.5\,\unit{\milli\radian}$) around the $\mathbf{x}$-axis to test whether the closed-loop modes enable maintaining a constant vertical force or displacement during 3\,mm-long tangential motion at speed of 0.1~\unit{\milli\meter}/\unit{\second} along the $\mathbf{y}$-direction, despite misaligned conditions.

\begin{figure}[htb!]
    \centering
    \includegraphics[width=0.99\columnwidth]{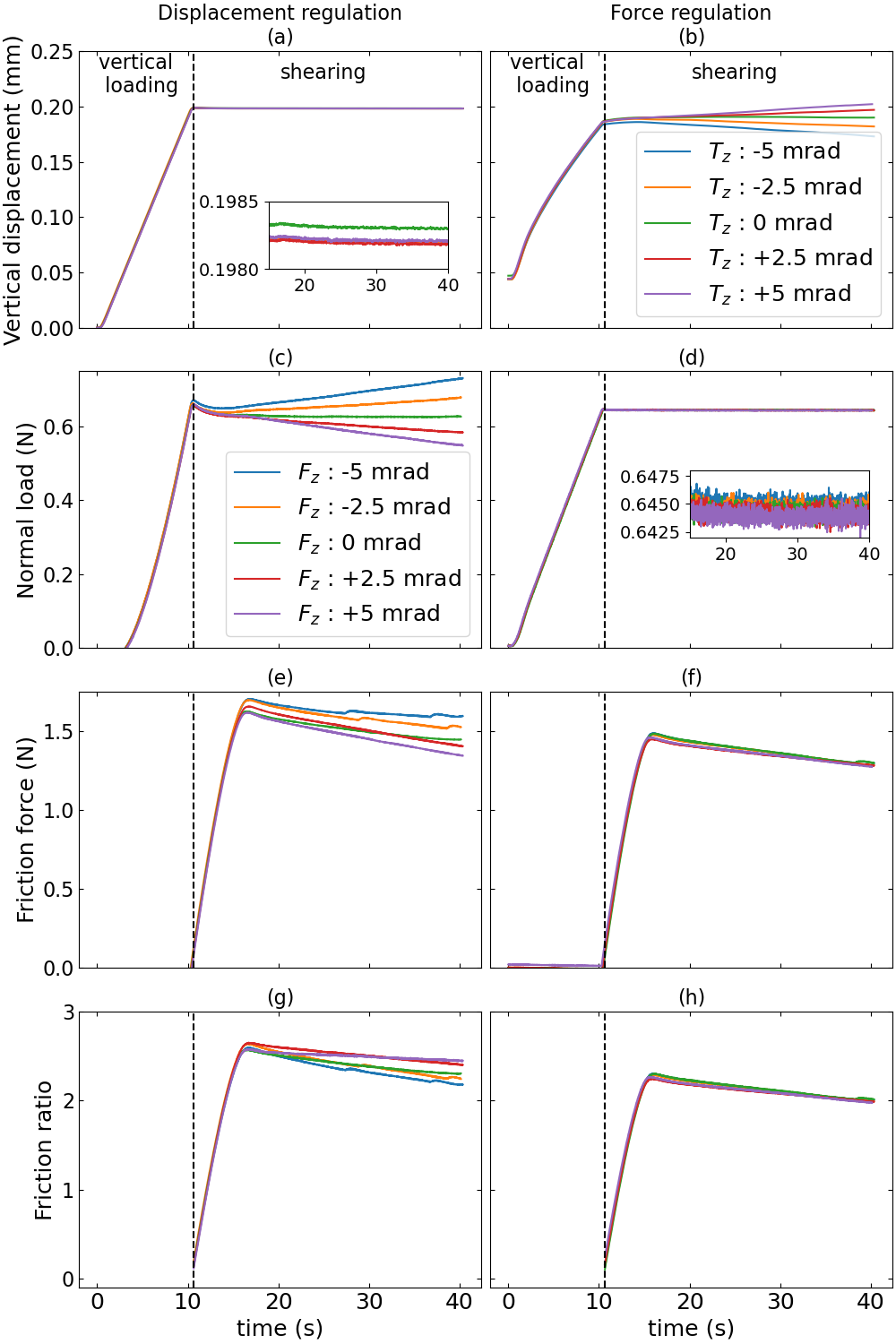}
    \caption{Friction experiments under imposed vertical force or displacement. An orientable glass plate is first pressed along $\mathbf{z}$ (normal loading phase) and then driven tangentially along $\mathbf{y}$ (shearing phase, starting at the vertical dashed line) on a PDMS sphere (Sylgard 184, curvature radius 9.42\,\unit{\milli\meter}). Left (resp. right) column: tests under imposed vertical displacement (resp. force) with set point 0.2\,\unit{\milli\meter} (resp. 0.6\,\unit{\newton}). The rows from top to bottom represent, respectively, the vertical position ($T_z$, \unit{\milli\meter}), normal force ($F_z$, \unit{\newton}), tangential force ($\sqrt{F_x^2+F_y^2}$, \unit{\newton}), and friction ratio ($\sqrt{F_x^2+F_y^2}/F_z$). Each plot displays five experiments with different tilt angles of the glass sample around $\mathbf{x}$, ranging from $-5$ to $5\,\unit{\milli\radian}$ (about 0.29\,\unit{\degree}) in steps of $2.5\,\unit{\milli\radian}$ [see legends in panels (b) and (c)]. The insets in panels (a) and (d) are zooms of the data in the corresponding main panel, illustrating the stability of the displacement and force regulation, respectively. Values of the friction ratio are only shown during the shearing phase.
    }
    \label{fig:multiple_friction}
\end{figure}

Figure~\ref{fig:multiple_friction} shows, for both control modes (left for displacement-controlled tests and right for force-controlled tests), the evolution of the vertical displacement, $T_z$, normal force, $F_z$, friction force, $\sqrt{F_x^2+F_y^2}$, and friction ratio ($\sqrt{F_x^2+F_y^2}/F_z$), across two phases: normal loading and tangential driving. The contact state reached after normal loading is very similar in both cases: the vertical displacement (resp. force) is about  0.2\,\unit{\milli\meter} (resp. 0.6\,\unit{\newton}).

In displacement-controlled mode, we first observe that the position regulation during sliding is nearly perfect irrespective of the tilt angle [see panel (a) of Fig.~\ref{fig:multiple_friction}]. This is achieved through tilt-dependent, continuous changes in the normal force [panel (c)], except in the test at $0\unit{\degree}$ where the force remains stable. These variations in normal force induce differences in friction force and friction ratio that depend on the tilt angle [panels (e) and (g)]. These results indicate that, in displacement-controlled experiments, the frictional state is very dependent on the quality of the alignment between the glass plate and the motion direction. \REV{For instance, for a misalignment of -2.5\,\unit{\milli\radian} (less than -0.15\,\unit{\degree}, a realistic value of sample mounting error; orange curves in Fig.~\ref{fig:multiple_friction}), the normalized root mean squared deviation (NRMSD) of the friction force [Fig.~\ref{fig:multiple_friction}(e)] with respect to the aligned case is more than 4.5\,\% under displacement control.} 

In force-controlled mode, it is now the normal force that remains very stable all along the sliding phase, regardless of the tilt angle [panel (d) of Fig.~\ref{fig:multiple_friction}].  
This is achieved through tilt-dependent continuous adjustment of the vertical position of the lower sample [panel (b)]. Importantly, we observe that neither the friction force nor the friction ratio depends on the misalignment [panels (f) and (h)]. \REV{In particular, for the orange curve shown in Fig.~\ref{fig:multiple_friction}(f), the NRMSD remains less than 0.63\,\% (to be compared to the above-mentioned 4.5\,\% under displacement control). Force control thus offers} the opportunity to conduct reproducible friction tests irrespective of any misalignment between the tangential motion and the glass plate. \REV{While misalignment-induced deviations in displacement-controlled conditions may be compensated by averaging the friction forces in the forward and reverse sliding in reciprocating tests on homogeneous contacts~\cite{burris_addressing_2009}, such compensation is not an option for, e.g., anisotropic samples~\cite{lu_novel_2018}, where a different value is expected in each of the two sliding directions.}

\subsection{Vibration-based detection of first contact}\label{sec:vibration}

Detecting the very first contact between two opaque approaching surfaces is a technological challenge in mechanical testing such as nano-indentation~\cite{bigerelle_first_2007,guillonneau_extraction_2012} or atomic force microscopy~\cite{gavara_combined_2016}. Even for nominally sharp indenters (e.g., pyramidal, like Berkovitch), actual indenters are often blunted, so that their tip can be approximated with a sphere. In this case, the force-displacement \REV{compression} curve initially shows sub-linear behavior, with a vanishing slope at first contact, making it difficult to detect above the unavoidable measurement noise.

\REV{The objective of the present illustration is to demonstrate the possibility of vibration-aided, on-the-fly detection of first contact. While careful post-treatment of a full compression experiment could help finding a posteriori the instant when first contact occurred~\cite{lucas_dynamics_1998}, here we target real-time detection, so that one may stop the approach movement as the contact is just created. Such capability is particularly interesting for hysteretic adhesive contacts, where growing and receding contacts follow different branches of the force--displacement curve. On-the-fly detection of first contact would for instance make it possible to create a zero-force (but finite area) contact on the compression branch, while late detection would require decreasing back the normal force to zero, following the other, separation branch.}

Our instrument is particularly suitable to implement such \REV{on-the-fly} vibration-aided first contact detection because (i) the piezoelectric table enables application of vibrations up to several tens of \unit{\hertz} and (ii) the lock-in function embedded into the FPGA enables detection of minute oscillations in the outputs of the force or displacement sensors in noisy environments. In this context, we carry out a \REV{macroscopic compression} experiment between a PDMS sphere with a curvature radius of 9.42~\unit{\milli\meter} fixed in the lower sample holder, and a glass plate fixed in the upper sample holder. The experiment is performed at a constant speed of \REV{0.4}\,\unit{\micro\meter}/\unit{\second} with the \REV{normal displacement} depth ramping in closed-loop control. \REV{In our methodology, the ground truth about the existence of a contact is provided by images of the contact,} taken simultaneously at a rate of 10\,\unit{\hertz}, from which the evolution of the real contact area, $A$, is extracted. The normal force, $F_\text{z}$, is measured along this progressive \REV{compression}.  During the experiment, a sinusoidal vibration \REV{normal to the contact ($\mathbf{z}$ axis) with amplitude of0.001 \unit{\volt} at 6\,\unit{\hertz} is also added to the output of the closed-loop function, resulting in a 168 \unit{\nano\meter} vertical oscillation of the piezoelectric table.} The lock-in detection \REV{(the chosen cutoff frequency and attenuation of the low-pass filter are 0.5\,\unit{\hertz} and 120\,\unit{\decibel}, leading to a response time of 1\,\unit{s})} is used to extract the amplitudes of the normal force and normal displacement components at 6\,\unit{\hertz}, $F_{\text{z,6Hz}}$ and $T_{\text{z,6Hz}}$ respectively. \REV{Note that the 6\,\unit{\hertz} frequency is sufficiently smaller than the lowest of the two resonance frequencies of the piezoelectric table (43\,\unit{\hertz}) and of the force/torque sensor (59\,\unit{\hertz}). Under these conditions, no vibration of the instrument is expected that could perturb the measurements made on the contact.}

\begin{figure}[h!]
    \centering
    \includegraphics[width=\columnwidth]{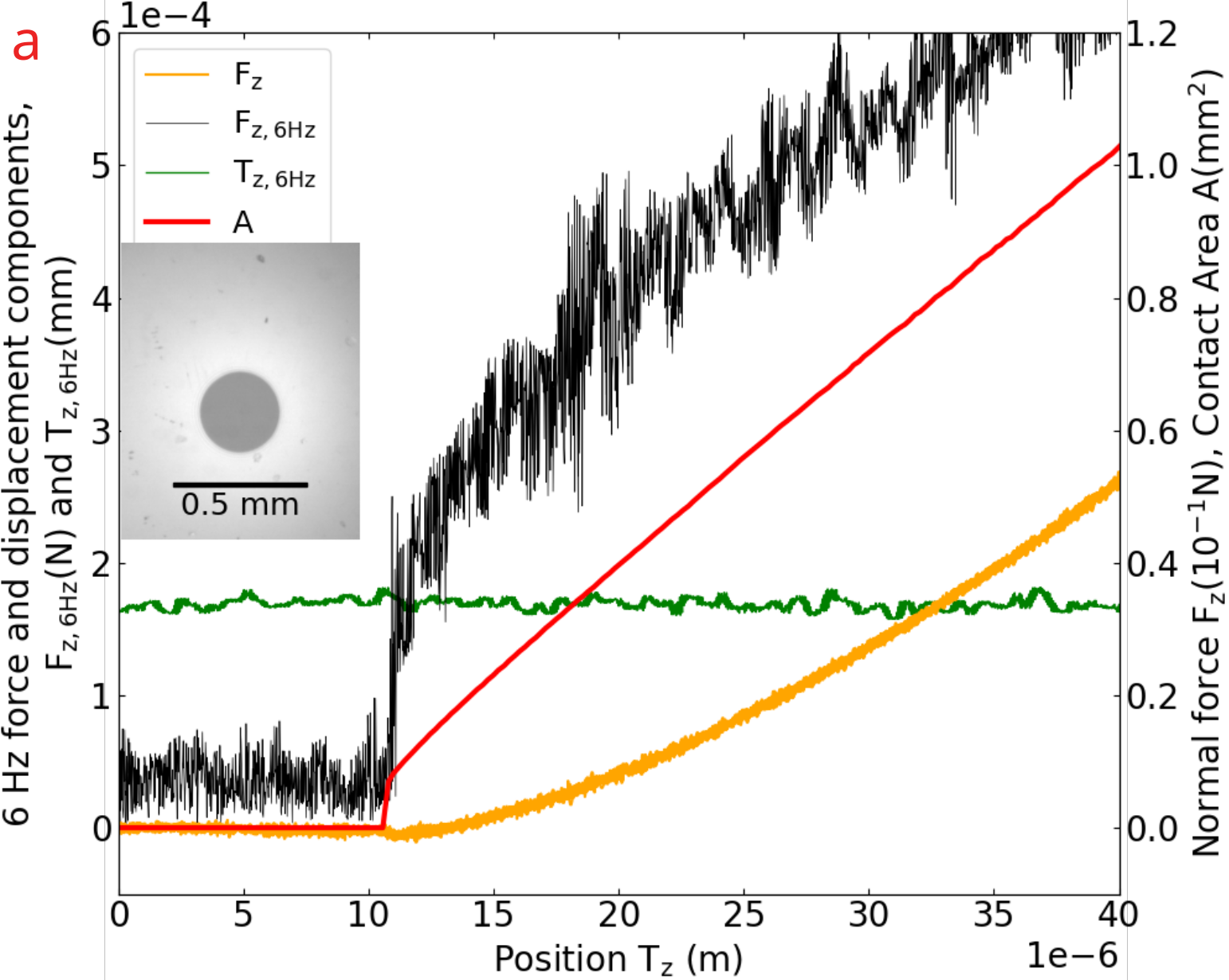}
    \includegraphics[width=\columnwidth]{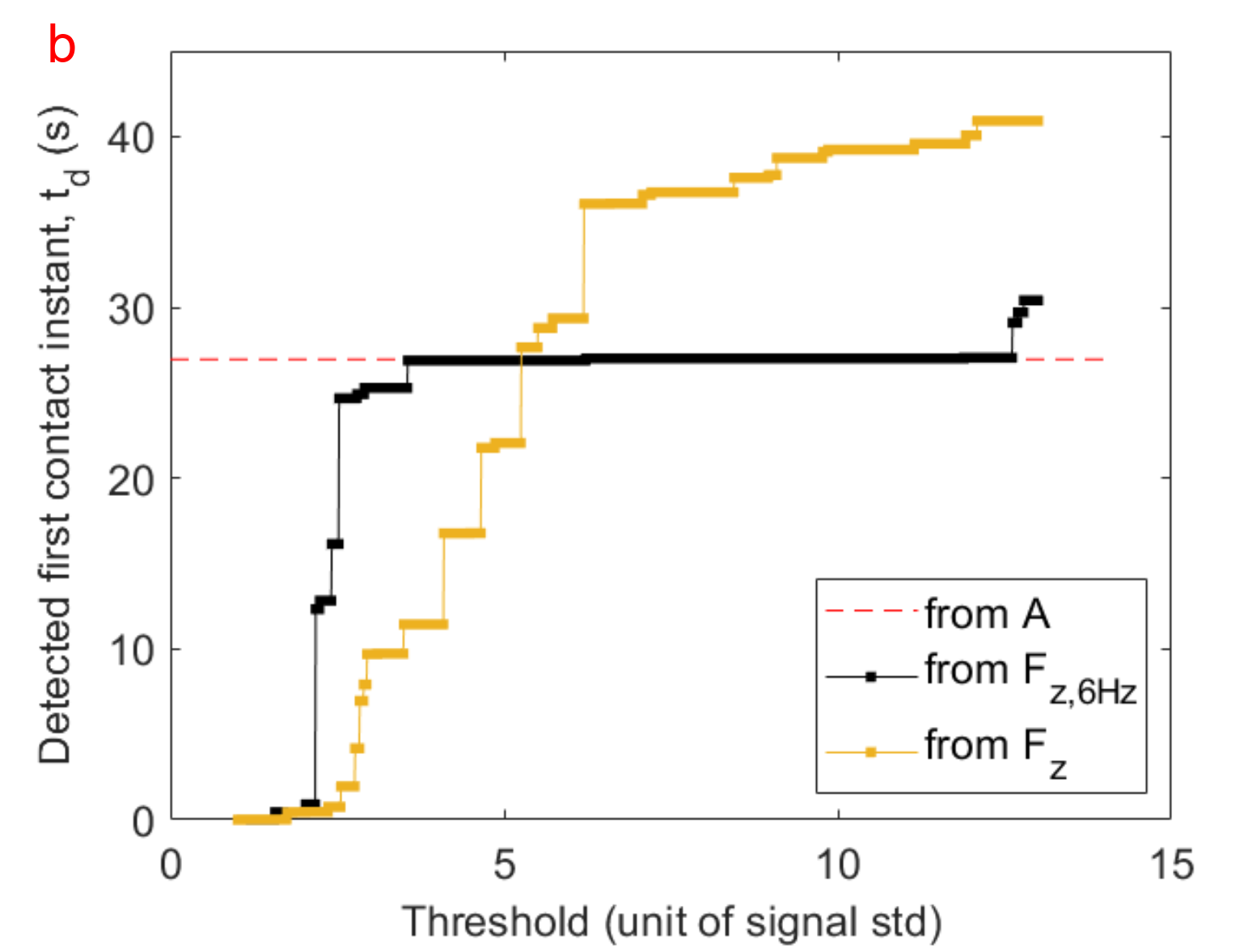}
    \caption{\REV{(a):} Evolutions of the normal force $F_\text{z}$ (orange curve), contact area $A$ (red) and lock-in--measured amplitudes of the 6\,\unit{\hertz}-components of the vertical displacement $T_{\text{z,6Hz}}$ (green) and force $F_{\text{z,6Hz}}$ (black) as functions of the vertical position, $T_\text{z}$, for a PDMS sphere with a radius of 9.42~\unit{\milli\meter} approaching and \REV{compressing} a glass plate. The sphere is submitted to an additional sinusoidal vibration (6\,\unit{\hertz} frequency, 168\,\unit{\nano\meter} amplitude) along the vertical axis. First contact, which occurs at around $T_\text{z}$=11\,\unit{\micro\meter}, \REV{causes an abrupt jump of} $F_{\text{z,6Hz}}$. The image of this first contact is represented in the inset. \REV{(b): Evolutions of the detected instant of first contact, $t_d$, for both the normal force signal ($F_\text{z}$, orange) and the lock-in response ($F_{\text{z,6Hz}}$, black) as a function of $b$, i.e., the threshold used for detection (in units of the standard deviation of the respective signal during the first 10\,\unit{\second}, when there is no contact). The red dashed line indicates the ground truth, provided by the detection of a non-vanishing contact area, $A$, in the images.}}
    \label{fig:first_contact}
\end{figure}

Figure~\ref{fig:first_contact}\REV{(a)} shows the evolutions of all measured quantities as a  function of the imposed vertical position, $T_\text{z}$, of the PDMS sphere during its approach and progressive \REV{compression} on the glass plate. As expected, the 6\,\unit{\hertz}-displacement component (green curve) remains constant at the amplitude of the imposed vibration (168\,\unit{\nano\meter}), independently of the contact state (out- or in-contact). In contrast, the 6\,\unit{\hertz}-force component (black) shows a clear separation between to phases: a first phase ($T_\text{z}<11$\,\unit{\micro\meter}) where the component remains close to zero, as expected before first contact, and a second phase ($T_\text{z}>11$\,\unit{\micro\meter}) where the component has a non-vanishing value well above the noise level in the first phase. The transition between the two phases is abrupt, and is interpreted as the occurrence of first contact. This interpretation is confirmed by looking at the evolution of the contact area (red): the area has a sharp jump to non-zero values exactly where $F_{\text{z,6Hz}}$ abruptly increases.

We emphasize that this method of detecting first contact is much more \REV{robust and} accurate than what could have been done using only the total force signal $F_\text{z}$ [orange curve shown in Fig.~\ref{fig:first_contact}\REV{(a)].
To demonstrate this, we apply a simple threshold-based detection to the time-evolutions of both $F_{\text{z}}$ and $F_{\text{z,6Hz}}$. For each quantity, we first consider the first 10\,\unit{\second} of the  reference out-of-contact signal and measure the corresponding mean value, $m$, and standard deviation, $\sigma$. We then measure the instant of first contact, $t_d$, as the instant when $\left|F-m\right|$ first exceeds $b \sigma$, with $b$ being an arbitrary value between 1 and 13. Fig.~\ref{fig:first_contact}(b) shows the evolution of $t_d$ as a function of $b$, for both $F_\text{z}$ (orange curve) and $F_\text{z,6Hz}$ (black curve). The ground truth, measured as the instant when the contact area in the images becomes non-vanishing, is denoted by the red dashed line. The lock-in--based detection shows a wide range of thresholds for which detection is accurate (between about 3.6 and 12.6). In contrast, for the detection based on the direct force measurement, there is no single threshold that can be used to correctly identify the first contact. This analysis demonstrates that the lock-in signal enables a much more robust and accurate on-the-fly detection of first contact than the raw force signal.}


These results demonstrate that minute vibration of the contact, coupled with a lock-in--based monitoring of the force component at the vibration frequency, is an efficient way of detecting the first contact between two solids without visualization. The sensitivity of such detection is comparable with that of contact imaging, but can potentially be accessed at a much faster rate than that of image-analysis--based contact identification and can be applied to non transparent solid interfaces as well.

\section{Conclusions}

The developed instrument enables actuating a solid sample with complex kinematics involving five degrees of freedom (three translations and two rotations) with nanometer/microradian resolution over strokes as large as 20\,\unit{\milli\meter} along $\mathbf{x}$ and $\mathbf{y}$, 0.9\,\unit{\milli\meter} along $\mathbf{z}$ and 20\,\unit{\degree} in $\theta_\text{X}$ and $\theta_\text{Y}$. High-frequency, small amplitude custom vibrations can be added to the motion along $\mathbf{z}$. The first solid can be brought into contact with another one with virtually no size limitation on the thickness of the solids. Their relative motion can be measured along 6-axes (three translations and three rotations) simultaneously, with \unit{\nano\meter} resolution over a range of a few \unit{\milli\meter}, while the three forces and three torques experienced at the contact can be measured simultaneously, respectively with 1\unit{\milli\newton} and 0.1\,\unit{\milli\newton}.\unit{\meter} resolution over ranges of 10\,\unit{\newton} and 0.5\,\unit{\newton}.\unit{\meter}. Note that the force/torque range can be modified by adjusting the stiffness of the blades that form the deformable part of the force/torque 6-axes sensor (see Ref.~\cite{guibert_versatile_2021} for details). The contact interface can be monitored optically \REV{throughout the tribological experiments} with spatial and temporal resolutions that depend on the specific camera. Closed-loop controls and lock-in detections can be applied to either the normal force or normal displacement.

So far, the instrument has been used to study adhesion, friction and contact mechanics in elastomer-based contact interfaces~\cite{oliver_adhesion_2023,acito_use_2023,aymard_designing_2024,fu_automated_2026,dancette_combining_2026}. \REV{In particular, the possibility to probe friction and adhesion simultaneously makes it ideal to investigate the elusive coupling between both phenomena, as done in Ref.~\cite{oliver_adhesion_2023}. More generally, the instrument} also has the potential to be leveraged as a versatile actuation, measurement and visualization platform to investigate other \REV{interfacial problems}, from adhesive tape peeling~\cite{dalbe_multiscale_2015} to solid-liquid adhesion~\cite{nagy_determination_2024}.

\begin{acknowledgments}
This work was supported by the LABEX iMUST (ANR-10-LABX-0064) of Université de Lyon, within the program "Investissements d'Avenir" (ANR-11-IDEX-0007) operated by the French National Research Agency (ANR). It was also supported by ANR through grants ANR-18-CE08-0011-01 (project PROMETAF), ANR-21-CE06-0048-01 (project WEEL) and ANR-22-CE06-0013-01 (project CLOSER). The authors are indebted to the Carnot institute Ingénierie@Lyon, labelled by the French National Research Agency (ANR), for its support and funding.
\end{acknowledgments}

\section*{Data availability}
The data that supports the findings of this study are available from the corresponding author upon reasonable request.

\section*{Credits}
\textbf{M. Guibert}: Conceptualization; Data Curation; Formal Analysis; Investigation; Methodology; Project Administration; Software; Supervision; Validation; Visualization; Writing/Original Draft Preparation. \textbf{A. Aymard}: Investigation; Validation. \textbf{C. Oliver}: Investigation; Validation. \textbf{T. Durand}: Resources. \textbf{B. Boulet}: Data Curation; Validation. \textbf{D. Gabriel Kashala}: Resources; Validation. \textbf{A. Mille}: Validation. \textbf{M. Leocmach}: Conceptualization. \textbf{L. Vanel}: Conceptualization; Funding Acquisition; Project Administration. \textbf{D. Dalmas}: Conceptualization; Supervision; Writing/Review \& Editing. \textbf{J. Scheibert}: Conceptualization; Funding Acquisition; Project Administration; Supervision; Writing/Original Draft Preparation.
\section{Bibliography}
\bibliographystyle{unsrtnat}
\bibliography{bibMISTRESS}

\begin{thebibliography}{68}
\providecommand{\natexlab}[1]{#1}
\providecommand{\url}[1]{\texttt{#1}}
\expandafter\ifx\csname urlstyle\endcsname\relax
  \providecommand{\doi}[1]{doi: #1}\else
  \providecommand{\doi}{doi: \begingroup \urlstyle{rm}\Url}\fi

\bibitem[Worobets and Wannop(2015)]{worobets_influence_2015}
Jay Worobets and John~William Wannop.
\newblock Influence of basketball shoe mass, outsole traction, and forefoot
  bending stiffness on three athletic movements.
\newblock \emph{Sports Biomechanics}, 14\penalty0 (3):\penalty0 351--360, 2015.
\newblock URL \url{https://doi.org/10.1080/14763141.2015.1084031}.

\bibitem[Tuononen(2014)]{tuononen_digital_2014}
Ari~J. Tuononen.
\newblock Digital {{Image Correlation}} to analyse stick--slip behaviour of
  tyre tread block.
\newblock \emph{Tribology International}, 69:\penalty0 70--76, January 2014.
\newblock URL
  \url{http://linkinghub.elsevier.com/retrieve/pii/S0301679X13003009}.

\bibitem[Scheibert et~al.(2009)Scheibert, Leurent, Prevost, and
  Debr{\'e}geas]{scheibert_role_2009}
J.~Scheibert, S.~Leurent, A.~Prevost, and G.~Debr{\'e}geas.
\newblock The {{Role}} of {{Fingerprints}} in the {{Coding}} of {{Tactile
  Information Probed}} with a {{Biomimetic Sensor}}.
\newblock \emph{Science}, 323\penalty0 (5920):\penalty0 1503--1506, March 2009.
\newblock URL \url{http://www.sciencemag.org/cgi/doi/10.1126/science.1166467}.

\bibitem[Spiers et~al.(2018)Spiers, Calli, and
  Dollar]{spiers_variablefriction_2018}
A.~J. Spiers, B.~Calli, and A.~M. Dollar.
\newblock Variable-{{Friction Finger Surfaces}} to {{Enable Within-Hand
  Manipulation}} via {{Gripping}} and {{Sliding}}.
\newblock \emph{IEEE Robotics and Automation Letters}, 3\penalty0 (4):\penalty0
  4116--4123, 2018.

\bibitem[{de Wijk} et~al.(2003){de Wijk}, Engelen, and Prinz]{dewijk_role_2003}
Ren{\'e}~A {de Wijk}, Lina Engelen, and Jon~F Prinz.
\newblock The role of intra-oral manipulation in the perception of sensory
  attributes.
\newblock \emph{Appetite}, 40\penalty0 (1):\penalty0 1--7, February 2003.
\newblock URL
  \url{http://www.sciencedirect.com/science/article/pii/S0195666302001721}.

\bibitem[Sahli et~al.(2020)Sahli, Prot, Wang, M{\"u}ser, Piovar{\v c}i, Didyk,
  and Bennewitz]{sahli_tactile_2020}
Riad Sahli, Aubin Prot, Anle Wang, Martin~H. M{\"u}ser, Michal Piovar{\v c}i,
  Piotr Didyk, and Roland Bennewitz.
\newblock Tactile perception of randomly rough surfaces.
\newblock \emph{Scientific Reports}, 10\penalty0 (1):\penalty0 15800, September
  2020.
\newblock URL \url{https://doi.org/10.1038/s41598-020-72890-y}.

\bibitem[Vakis et~al.(2018)Vakis, Yastrebov, Scheibert, Nicola, Dini, Minfray,
  Almqvist, Paggi, Lee, Limbert, Molinari, Anciaux, Aghababaei, Restrepo,
  Papangelo, Cammarata, Nicolini, Putignano, Carbone, Stupkiewicz, Lengiewicz,
  Costagliola, Bosia, Guarino, Pugno, M{\"u}ser, and
  Ciavarella]{vakis_modeling_2018}
A.~I. Vakis, V.~A. Yastrebov, J.~Scheibert, L.~Nicola, D.~Dini, C.~Minfray,
  A.~Almqvist, M.~Paggi, S.~Lee, G.~Limbert, J.~F. Molinari, G.~Anciaux,
  R.~Aghababaei, S.~Echeverri Restrepo, A.~Papangelo, A.~Cammarata,
  P.~Nicolini, C.~Putignano, G.~Carbone, S.~Stupkiewicz, J.~Lengiewicz,
  G.~Costagliola, F.~Bosia, R.~Guarino, N.~M. Pugno, M.~H. M{\"u}ser, and
  M.~Ciavarella.
\newblock Modeling and simulation in tribology across scales: {{An}} overview.
\newblock \emph{Tribology International}, 125:\penalty0 169--199, 2018.
\newblock URL \url{https://doi.org/10.1016/j.triboint.2018.02.005}.

\bibitem[Weber et~al.(2022)Weber, Scheibert, {de Boer}, and
  Dhinojwala]{weber_experimental_2022}
Bart Weber, Julien Scheibert, Maarten~P. {de Boer}, and Ali Dhinojwala.
\newblock Experimental insights into adhesion and friction between nominally
  dry rough surfaces.
\newblock \emph{MRS Bulletin}, 47\penalty0 (12):\penalty0 1237--1246, December
  2022.
\newblock URL \url{https://doi.org/10.1557/s43577-022-00464-6}.

\bibitem[Scheibert et~al.(2027)Scheibert, Bayart, Bonn, {Cayer-Barrioz},
  {Dwyer-Joyce}, and Mazuyer]{scheibert_measuring_2027}
Julien Scheibert, Elsa Bayart, Daniel Bonn, Juliette {Cayer-Barrioz}, Robert
  {Dwyer-Joyce}, and Denis Mazuyer.
\newblock Measuring {{Mechanical Fields}} in {{Tribology}}: {{Physical
  Quantities}}, {{Methods}}, and {{Insights}}.
\newblock In Jun Qu and Stefanie Hanke, editors, \emph{Encyclopedia of
  {{Tribology}} and {{Lubrication}} ({{First Edition}})}, volume~3, pages
  3--24. Elsevier, London, first edition, 2027.
\newblock ISBN 978-0-443-30139-1.
\newblock URL
  \url{https://www.sciencedirect.com/science/article/pii/B9780443301384000315}.

\bibitem[M{\"u}ser et~al.(2017)M{\"u}ser, Dapp, Bugnicourt, Sainsot, Lesaffre,
  Lubrecht, Persson, Harris, Bennett, Schulze, Rohde, Ifju, Sawyer, Angelini,
  Ashtari~Esfahani, Kadkhodaei, Akbarzadeh, Wu, Vorlaufer, Vernes, Solhjoo,
  Vakis, Jackson, Xu, Streator, Rostami, Dini, Medina, Carbone, Bottiglione,
  Afferrante, Monti, Pastewka, Robbins, and Greenwood]{muser_meeting_2017}
Martin~H. M{\"u}ser, Wolf~B. Dapp, Romain Bugnicourt, Philippe Sainsot, Nicolas
  Lesaffre, Ton~A. Lubrecht, Bo~N.~J. Persson, Kathryn Harris, Alexander
  Bennett, Kyle Schulze, Sean Rohde, Peter Ifju, W.~Gregory Sawyer, Thomas
  Angelini, Hossein Ashtari~Esfahani, Mahmoud Kadkhodaei, Saleh Akbarzadeh,
  Jiunn-Jong Wu, Georg Vorlaufer, Andr{\'a}s Vernes, Soheil Solhjoo, Antonis~I.
  Vakis, Robert~L. Jackson, Yang Xu, Jeffrey Streator, Amir Rostami, Daniele
  Dini, Simon Medina, Giuseppe Carbone, Francesco Bottiglione, Luciano
  Afferrante, Joseph Monti, Lars Pastewka, Mark~O. Robbins, and James~A.
  Greenwood.
\newblock Meeting the {{Contact-Mechanics Challenge}}.
\newblock \emph{Tribology Letters}, 65:\penalty0 118, December 2017.
\newblock URL \url{http://link.springer.com/10.1007/s11249-017-0900-2}.

\bibitem[Sahli et~al.(2018)Sahli, Pallares, Ducottet, Ben~Ali, Al~Akhrass,
  Guibert, and Scheibert]{sahli_evolution_2018}
Riad Sahli, Ga{\"e}l Pallares, Christophe Ducottet, Imed~Eddine Ben~Ali, Samer
  Al~Akhrass, Matthieu Guibert, and Julien Scheibert.
\newblock Evolution of real contact area under shear and the value of static
  friction of soft materials.
\newblock \emph{Proceedings of the National Academy of Sciences of the USA},
  115:\penalty0 471--176, 2018.
\newblock URL \url{https://doi.org/10.1073/pnas.1706434115}.

\bibitem[Baumberger et~al.(2003)Baumberger, Caroli, and
  Ronsin]{baumberger_selfhealing_2003}
T.~Baumberger, C.~Caroli, and O.~Ronsin.
\newblock Self-healing slip pulses and the friction of gelatin gels.
\newblock \emph{The European Physical Journal E - Soft Matter}, 11\penalty0
  (1):\penalty0 85--93, May 2003.
\newblock URL
  \url{http://www.springerlink.com/Index/10.1140/epje/i2003-10009-7}.

\bibitem[Nase et~al.(2008)Nase, Lindner, and Creton]{nase_pattern_2008}
Julia Nase, Anke Lindner, and Costantino Creton.
\newblock Pattern {{Formation}} during {{Deformation}} of a {{Confined
  Viscoelastic Layer}}: {{From}} a {{Viscous Liquid}} to a {{Soft Elastic
  Solid}}.
\newblock \emph{Physical Review Letters}, 101\penalty0 (7):\penalty0 074503,
  August 2008.
\newblock URL \url{https://link.aps.org/doi/10.1103/PhysRevLett.101.074503}.

\bibitem[Prevost et~al.(2013)Prevost, Scheibert, and
  Debr{\'e}geas]{prevost_probing_2013}
A.~Prevost, J.~Scheibert, and G.~Debr{\'e}geas.
\newblock Probing the micromechanics of a multi-contact interface at the onset
  of frictional sliding.
\newblock \emph{The European Physical Journal E}, 36\penalty0 (2), February
  2013.
\newblock URL \url{http://link.springer.com/10.1140/epje/i2013-13017-0}.

\bibitem[Delhaye et~al.(2014)Delhaye, Lefevre, and
  Thonnard]{delhaye_dynamics_2014}
Benoit Delhaye, Philippe Lefevre, and Jean-Louis Thonnard.
\newblock Dynamics of fingertip contact during the onset of tangential slip.
\newblock \emph{Journal of the Royal Society Interface}, 11\penalty0
  (100):\penalty0 20140698, November 2014.

\bibitem[Dalbe et~al.(2015)Dalbe, Cortet, Ciccotti, Vanel, and
  Santucci]{dalbe_multiscale_2015}
Marie-Julie Dalbe, Pierre-Philippe Cortet, Matteo Ciccotti, Lo{\"i}c Vanel, and
  St{\'e}phane Santucci.
\newblock Multiscale {{Stick-Slip Dynamics}} of {{Adhesive Tape Peeling}}.
\newblock \emph{Physical Review Letters}, 115\penalty0 (12), September 2015.
\newblock URL \url{https://link.aps.org/doi/10.1103/PhysRevLett.115.128301}.

\bibitem[McGhee et~al.(2017)McGhee, Pitenis, Bennett, Harris, Schulze,
  Urue{\~n}a, Ifju, Angelini, M{\"u}ser, and Sawyer]{mcghee_contact_2017}
Alexander~J. McGhee, Angela~A. Pitenis, Alexander~I. Bennett, Kathryn~L.
  Harris, Kyle~D. Schulze, Juan~Manuel Urue{\~n}a, Peter~G. Ifju, Thomas~E.
  Angelini, Martin~H. M{\"u}ser, and W.~Gregory Sawyer.
\newblock Contact and {{Deformation}} of {{Randomly Rough Surfaces}} with
  {{Varying Root-Mean-Square Gradient}}.
\newblock \emph{Tribology Letters}, 65\penalty0 (4):\penalty0 157, November
  2017.
\newblock URL \url{https://doi.org/10.1007/s11249-017-0942-5}.

\bibitem[Sahli et~al.(2019)Sahli, Pallares, Papangelo, Ciavarella, Ducottet,
  Ponthus, and Scheibert]{sahli_shearinduced_2019}
R.~Sahli, G.~Pallares, A.~Papangelo, M.~Ciavarella, C.~Ducottet, N.~Ponthus,
  and J.~Scheibert.
\newblock Shear-{{Induced Anisotropy}} in {{Rough Elastomer Contact}}.
\newblock \emph{Physical Review Letters}, 122:\penalty0 214301, May 2019.
\newblock URL \url{https://doi.org/10.1103/PhysRevLett.122.214301}.

\bibitem[Scheibert et~al.(2020)Scheibert, Sahli, and
  Peyrard]{scheibert_onset_2020}
Julien Scheibert, Riad Sahli, and Michel Peyrard.
\newblock Onset of {{Sliding}} of {{Elastomer Multicontacts}}: {{Failure}} of a
  {{Model}} of {{Independent Asperities}} to {{Match Experiments}}.
\newblock \emph{Frontiers in Mechanical Engineering}, 6:\penalty0 18, 2020.
\newblock URL
  \url{https://www.frontiersin.org/article/10.3389/fmech.2020.00018}.

\bibitem[Xu et~al.(2022)Xu, Scheibert, Gadegaard, and
  Mulvihill]{xu_asperitybased_2022}
Yang Xu, Julien Scheibert, Nikolaj Gadegaard, and Daniel~M. Mulvihill.
\newblock An asperity-based statistical model for the adhesive friction of
  elastic nominally flat rough contact interfaces.
\newblock \emph{Journal of the Mechanics and Physics of Solids}, 164:\penalty0
  104878, July 2022.
\newblock URL
  \url{https://www.sciencedirect.com/science/article/pii/S0022509622000886}.

\bibitem[Putignano et~al.(2013)Putignano, Reddyhoff, Carbone, and
  Dini]{putignano_experimental_2013}
Carmine Putignano, Thomas Reddyhoff, Giuseppe Carbone, and Daniele Dini.
\newblock Experimental {{Investigation}} of {{Viscoelastic Rolling Contacts}}:
  {{A Comparison}} with {{Theory}}.
\newblock \emph{Tribology Letters}, 51\penalty0 (1):\penalty0 105--113, July
  2013.
\newblock URL \url{https://doi.org/10.1007/s11249-013-0151-9}.

\bibitem[Scaraggi and Persson(2015)]{scaraggi_friction_2015}
M~Scaraggi and B~N~J Persson.
\newblock Friction and universal contact area law for randomly rough
  viscoelastic contacts.
\newblock \emph{Journal of Physics: Condensed Matter}, 27\penalty0
  (10):\penalty0 105102, March 2015.
\newblock URL
  \url{http://stacks.iop.org/0953-8984/27/i=10/a=105102?key=crossref.8004c5a86aac96cae2c24be37d49378e}.

\bibitem[Shoaib and {Espinosa-Marzal}(2019)]{shoaib_influence_2019}
Tooba Shoaib and Rosa~M. {Espinosa-Marzal}.
\newblock Influence of {{Loading Conditions}} and {{Temperature}} on {{Static
  Friction}} and {{Contact Aging}} of {{Hydrogels}} with {{Modulated
  Microstructures}}.
\newblock \emph{ACS Applied Materials \& Interfaces}, 11\penalty0
  (45):\penalty0 42722--42733, November 2019.
\newblock URL \url{https://doi.org/10.1021/acsami.9b14283}.

\bibitem[Yin et~al.(2004)Yin, Ling, and Liu]{yin_dynamic_2004}
Yanling Yin, Shih-Fu Ling, and Yong Liu.
\newblock A dynamic indentation method for characterizing soft incompressible
  viscoelastic materials.
\newblock \emph{Materials Science and Engineering: A}, 379\penalty0
  (1):\penalty0 334--340, August 2004.
\newblock URL
  \url{http://www.sciencedirect.com/science/article/pii/S0921509304002321}.

\bibitem[Boyer et~al.(2009)Boyer, Laqui{\`e}ze, Le~Bot, Laqui{\`e}ze, and
  Zahouani]{boyer_dynamic_2009}
G.~Boyer, L.~Laqui{\`e}ze, A.~Le~Bot, S.~Laqui{\`e}ze, and H.~Zahouani.
\newblock Dynamic indentation on human skin in vivo: Ageing effects.
\newblock \emph{Skin Research and Technology}, 15\penalty0 (1):\penalty0
  55--67, February 2009.
\newblock URL \url{https://doi.org/10.1111/j.1600-0846.2008.00324.x}.

\bibitem[{Samadi-Dooki} et~al.(2017){Samadi-Dooki}, Voyiadjis, and
  Stout]{samadi-dooki_indirect_2017}
Aref {Samadi-Dooki}, George~Z. Voyiadjis, and Rhett~W. Stout.
\newblock An {{Indirect Indentation Method}} for {{Evaluating}} the {{Linear
  Viscoelastic Properties}} of the {{Brain Tissue}}.
\newblock \emph{Journal of Biomechanical Engineering}, 139\penalty0 (061007),
  April 2017.
\newblock URL \url{https://doi.org/10.1115/1.4036486}.

\bibitem[Wahl et~al.(2006)Wahl, Asif, Greenwood, and
  Johnson]{wahl_oscillating_2006}
K.J. Wahl, S.A.S. Asif, J.A. Greenwood, and K.L. Johnson.
\newblock Oscillating adhesive contacts between micron-scale tips and compliant
  polymers.
\newblock \emph{Journal of Colloid and Interface Science}, 296\penalty0
  (1):\penalty0 178--188, April 2006.
\newblock URL
  \url{http://www.sciencedirect.com/science/article/pii/S0021979705008787}.

\bibitem[Charrault et~al.(2009)Charrault, Gauthier, Marie, and
  Schirrer]{charrault_experimental_2009}
E.~Charrault, C.~Gauthier, P.~Marie, and R.~Schirrer.
\newblock Experimental and {{Theoretical Analysis}} of a {{Dynamic JKR
  Contact}}.
\newblock \emph{Langmuir}, 25\penalty0 (10):\penalty0 5847--5854, May 2009.
\newblock URL \url{https://doi.org/10.1021/la803434b}.

\bibitem[Tricarico et~al.(2025)Tricarico, Ciavarella, and
  Papangelo]{tricarico_enhancement_2025}
Michele Tricarico, Michele Ciavarella, and Antonio Papangelo.
\newblock Enhancement of adhesion strength through microvibrations:
  {{Modeling}} and experiments.
\newblock \emph{Journal of the Mechanics and Physics of Solids}, 196:\penalty0
  106020, 2025.
\newblock URL
  \url{https://www.sciencedirect.com/science/article/pii/S0022509624004861}.

\bibitem[Yi et~al.(2025)Yi, Haouas, Gauthier, and Rabenorosoa]{yi_pdms_2025}
Jiawei Yi, Wissem Haouas, Micha{\"e}l Gauthier, and Kanty Rabenorosoa.
\newblock A {{PDMS}}/{{Silicon Adhesion Control Method}} at {{Millimeter-Scale
  Based}} on {{Microvibration}}.
\newblock \emph{Advanced Intelligent Systems}, 7\penalty0 (2):\penalty0
  2400394, 2025.
\newblock URL
  \url{https://advanced.onlinelibrary.wiley.com/doi/abs/10.1002/aisy.202400394}.

\bibitem[Tricarico et~al.(2026)Tricarico, Shiferaw, and
  Papangelo]{tricarico_influence_2026}
M.~Tricarico, A.~Y. Shiferaw, and A.~Papangelo.
\newblock Influence of material and geometrical parameters on the adhesive
  performance of vibration-modulated soft contacts.
\newblock \emph{European Journal of Mechanics - A/Solids}, 118:\penalty0
  106130, 2026.
\newblock URL
  \url{https://www.sciencedirect.com/science/article/pii/S0997753826001166}.

\bibitem[Chen et~al.(2026)Chen, Chen, Zhao, Jiang, Zhu, Xu, Zhao, Tang, Gorb,
  Ji, and Dai]{chen_ultrafast_2026}
Jian Chen, Wenjie Chen, Mengmeng Zhao, Xingyu Jiang, Xiaolei Zhu, Xinran Xu,
  Jiahui Zhao, Yezhong Tang, Stanislav~N Gorb, Keju Ji, and Zhendong Dai.
\newblock Ultrafast {{Adhesion}}/{{Friction Bidirectionally Switchable
  Control}} by {{Vibration}}.
\newblock \emph{Advanced Functional Materials}, 36\penalty0 (8):\penalty0
  e16421, 2026.
\newblock URL
  \url{https://advanced.onlinelibrary.wiley.com/doi/abs/10.1002/adfm.202516421}.

\bibitem[Carpick et~al.(1997)Carpick, Ogletree, and
  Salmeron]{carpick_lateral_1997}
R.~W. Carpick, D.~F. Ogletree, and M.~Salmeron.
\newblock Lateral stiffness: {{A}} new nanomechanical measurement for the
  determination of shear strengths with friction force microscopy.
\newblock \emph{Applied Physics Letters}, 70\penalty0 (12):\penalty0
  1548--1550, March 1997.
\newblock URL \url{https://doi.org/10.1063/1.118639}.

\bibitem[Benad et~al.(2019)Benad, Nakano, Popov, and Popov]{benad_active_2019}
J.~Benad, K.~Nakano, V.~L. Popov, and M.~Popov.
\newblock Active control of friction by transverse oscillations.
\newblock \emph{Friction}, 7\penalty0 (1):\penalty0 74--85, February 2019.
\newblock URL \url{https://doi.org/10.1007/s40544-018-0202-1}.

\bibitem[Dorogin et~al.(2017)Dorogin, Tiwari, Rotella, Mangiagalli, and
  Persson]{dorogin_role_2017}
L.~Dorogin, A.~Tiwari, C.~Rotella, P.~Mangiagalli, and B.~N.~J. Persson.
\newblock Role of {{Preload}} in {{Adhesion}} of {{Rough Surfaces}}.
\newblock \emph{Physical Review Letters}, 118\penalty0 (23):\penalty0 238001,
  June 2017.
\newblock URL \url{http://link.aps.org/doi/10.1103/PhysRevLett.118.238001}.

\bibitem[Aleshin and Bou~Matar(2016)]{aleshin_solution_2016}
V.~V. Aleshin and O.~Bou~Matar.
\newblock Solution to the frictional contact problem via the method of memory
  diagrams for general {{3D}} loading histories.
\newblock \emph{Physical Mesomechanics}, 19\penalty0 (2):\penalty0 130--135,
  April 2016.
\newblock URL \url{https://doi.org/10.1134/S102995991602003X}.

\bibitem[Popov et~al.(2015)Popov, Popov, and Pohrt]{popov_relaxation_2015}
M.~Popov, V.L. Popov, and R.~Pohrt.
\newblock Relaxation damping in oscillating contacts.
\newblock \emph{Scientific Reports}, 5\penalty0 (1), December 2015.
\newblock URL \url{http://www.nature.com/articles/srep16189}.

\bibitem[Man{\'e} et~al.(2013)Man{\'e}, Loubet, Guerret, Guy, Sanseau, Odoni,
  Vanel, Long, and Sotta]{mane_new_2013}
Z.~Man{\'e}, J.-L. Loubet, C.~Guerret, L.~Guy, O.~Sanseau, L.~Odoni, L.~Vanel,
  D.R. Long, and P.~Sotta.
\newblock A new rotary tribometer to study the wear of reinforced rubber
  materials.
\newblock \emph{4th UK-China Tribology}, 306\penalty0 (1):\penalty0 149--160,
  August 2013.
\newblock URL
  \url{http://www.sciencedirect.com/science/article/pii/S0043164813004420}.

\bibitem[Chanal et~al.(2025)Chanal, Galipaud, Moreaux, Loubet, and
  Sotta]{chanal_characterization_2025}
C.~Chanal, J.~Galipaud, B.~Moreaux, J.-L. Loubet, and P.~Sotta.
\newblock Characterization of oxidative processes associated to low-severity
  tire tread wear.
\newblock \emph{Wear}, 566--567:\penalty0 205753, 2025.
\newblock URL
  \url{https://www.sciencedirect.com/science/article/pii/S0043164825000225}.

\bibitem[Huang et~al.(2018)Huang, Guibert, Th{\'e}venet, Fayolle, Chauss{\'e}e,
  Guy, Vanel, Loubet, and Sotta]{huang_new_2018}
M.~Huang, M.~Guibert, J.~Th{\'e}venet, C.~Fayolle, T.~Chauss{\'e}e, L.~Guy,
  L.~Vanel, J.-L. Loubet, and P.~Sotta.
\newblock A new test method to simulate low-severity wear conditions
  experienced by rubber tire materials.
\newblock \emph{Wear}, 410--411:\penalty0 72--82, 2018.
\newblock URL
  \url{https://www.sciencedirect.com/science/article/pii/S0043164818302771}.

\bibitem[Gravish et~al.(2008)Gravish, Wilkinson, and
  Autumn]{gravish_frictional_2008}
Nick Gravish, Matt Wilkinson, and Kellar Autumn.
\newblock Frictional and elastic energy in gecko adhesive detachment.
\newblock \emph{Journal of The Royal Society Interface}, 5\penalty0
  (20):\penalty0 339--348, March 2008.
\newblock URL \url{https://doi.org/10.1098/rsif.2007.1077}.

\bibitem[Ha et~al.(2025)Ha, Yoo, Li, Mao, Xu, Qi, Wu, Fan, Yuan, Kim, Flavin,
  Yoo, Shahir, Kim, Ahn, Colgate, Huang, and Rogers]{ha_full_2025}
Kyoung-Ho Ha, Jaeyoung Yoo, Shupeng Li, Yuxuan Mao, Shengwei Xu, Hongyuan Qi,
  Hanbing Wu, Chengye Fan, Hanyin Yuan, Jin-Tae Kim, Matthew~T. Flavin,
  Seonggwang Yoo, Pratyush Shahir, Sangjun Kim, Hak-Young Ahn, Edward Colgate,
  Yonggang Huang, and John~A. Rogers.
\newblock Full freedom-of-motion actuators as advanced haptic interfaces.
\newblock \emph{Science}, 387\penalty0 (6741):\penalty0 1383--1390, 2025.
\newblock URL \url{https://www.science.org/doi/abs/10.1126/science.adt2481}.

\bibitem[Mergel et~al.(2019)Mergel, Sahli, Scheibert, and
  Sauer]{mergel_continuum_2019}
Janine~C. Mergel, Riad Sahli, Julien Scheibert, and Roger~A. Sauer.
\newblock Continuum contact models for coupled adhesion and friction.
\newblock \emph{The Journal of Adhesion}, 95\penalty0 (12):\penalty0
  1101--1133, 2019.
\newblock URL \url{https://doi.org/10.1080/00218464.2018.1479258}.

\bibitem[Guibert et~al.(2021)Guibert, Oliver, Durand, Le~Mogne, Le~Bot, Dalmas,
  Scheibert, and Fontaine]{guibert_versatile_2021}
M.~Guibert, C.~Oliver, T.~Durand, T.~Le~Mogne, A.~Le~Bot, D.~Dalmas,
  J.~Scheibert, and J.~Fontaine.
\newblock A versatile flexure-based six-axis force/torque sensor and its
  application to tribology.
\newblock \emph{Review of Scientific Instruments}, 92\penalty0 (8):\penalty0
  085002, 2021.
\newblock URL \url{https://doi.org/10.1063/5.0057266}.

\bibitem[Stewart(1965)]{stewart_platform_1965}
D.~Stewart.
\newblock A {{Platform}} with {{Six Degrees}} of {{Freedom}}.
\newblock \emph{Proceedings of the Institution of Mechanical Engineers},
  180\penalty0 (1):\penalty0 371--386, 1965.
\newblock URL \url{https://doi.org/10.1243/PIME_PROC_1965_180_029_02}.

\bibitem[Meade(1982)]{meade_advances_1982}
M.~L. Meade.
\newblock Advances in lock-in amplifiers.
\newblock \emph{Journal of Physics E: Scientific Instruments}, 15\penalty0
  (4):\penalty0 395, April 1982.
\newblock URL \url{https://doi.org/10.1088/0022-3735/15/4/001}.

\bibitem[Zhang et~al.(2024{\natexlab{a}})Zhang, Jeong, and
  Kang]{zhang_lockin_2024}
Qianwen Zhang, Wonje Jeong, and Dae~Joon Kang.
\newblock Lock-in amplifiers as a platform for weak signal measurements:
  {{Development}} and applications.
\newblock \emph{Current Applied Physics}, 66:\penalty0 95--109,
  2024{\natexlab{a}}.
\newblock URL
  \url{https://www.sciencedirect.com/science/article/pii/S1567173924001512}.

\bibitem[{Galaviz-Aguilar} et~al.(2025){Galaviz-Aguilar}, {Vargas-Rosales},
  Falcone, and {Aguilar-Avelar}]{galaviz-aguilar_fieldprogrammable_2025}
Jose~Alejandro {Galaviz-Aguilar}, Cesar {Vargas-Rosales}, Francisco Falcone,
  and Carlos {Aguilar-Avelar}.
\newblock Field-{{Programmable Gate Array}} ({{FPGA}})-{{Based Lock-In
  Amplifier System}} with {{Signal Enhancement}}: {{A Comprehensive Review}} on
  the {{Design}} for {{Advanced Measurement Applications}}.
\newblock \emph{Sensors}, 25\penalty0 (2), 2025.
\newblock URL \url{https://www.mdpi.com/1424-8220/25/2/584}.

\bibitem[Hale(1999)]{hale_principles_1999}
L.~C. Hale.
\newblock \emph{Principles and Techniques for Designing Precision Machines}.
\newblock PhD thesis, University of California, 1999.
\newblock URL \url{https://www.osti.gov/biblio/8431}.

\bibitem[Aymard et~al.(2024)Aymard, Delplanque, Dalmas, and
  Scheibert]{aymard_designing_2024}
A.~Aymard, E.~Delplanque, D.~Dalmas, and J.~Scheibert.
\newblock Designing metainterfaces with specified friction laws.
\newblock \emph{Science}, 383\penalty0 (6679):\penalty0 200--204, 2024.

\bibitem[Zeka et~al.(2026)Zeka, Blal, Fekak, Duval, Gravouil, and
  Scheibert]{zeka_normal_2026}
Donald Zeka, Nawfal Blal, Fatima-Ezzahra Fekak, Arnaud Duval, Anthony Gravouil,
  and Julien Scheibert.
\newblock Normal contact of metainterfaces: {{The}} roles of finite size and
  microcontact interactions.
\newblock \emph{Journal of the Mechanics and Physics of Solids}, 214:\penalty0
  106646, 2026.
\newblock URL
  \url{https://www.sciencedirect.com/science/article/pii/S0022509626001468}.

\bibitem[Fu et~al.(2026)Fu, Gabriel~Kashala, Dalmas, and
  Scheibert]{fu_automated_2026}
Li~Fu, Djibril Gabriel~Kashala, Davy Dalmas, and Julien Scheibert.
\newblock Automated inverse design of metainterfaces with tailored friction
  laws.
\newblock \emph{Nature Physics}, October 2026.
\newblock URL \url{https://doi.org/10.1038/s41567-026-03503-0}.

\bibitem[Scheibert(2027)]{scheibert_slope_2027}
Julien Scheibert.
\newblock The slope of the friction law of hertzian-asperity-based
  metainterfaces has a finite positive lower bound.
\newblock \emph{Tribology International}, 225:\penalty0 112422, 2027.
\newblock URL
  \url{https://www.sciencedirect.com/science/article/pii/S0301679X26007656}.

\bibitem[Papangelo et~al.(2019)Papangelo, Scheibert, Sahli, Pallares, and
  Ciavarella]{papangelo_shearinduced_2019}
A.~Papangelo, J.~Scheibert, R.~Sahli, G.~Pallares, and M.~Ciavarella.
\newblock Shear-induced contact area anisotropy explained by a fracture
  mechanics model.
\newblock \emph{Physical Review E}, 99\penalty0 (5):\penalty0 053005, May 2019.
\newblock URL \url{https://doi.org/10.1103/PhysRevE.99.053005}.

\bibitem[Lengiewicz et~al.(2020)Lengiewicz, {de Souza}, Lahmar, Courbon,
  Dalmas, Stupkiewicz, and Scheibert]{lengiewicz_finite_2020}
J.~Lengiewicz, M.~{de Souza}, M.~A. Lahmar, C.~Courbon, D.~Dalmas,
  S.~Stupkiewicz, and J.~Scheibert.
\newblock Finite deformations govern the anisotropic shear-induced area
  reduction of soft elastic contacts.
\newblock \emph{Journal of the Mechanics and Physics of Solids}, 143:\penalty0
  104056, 2020.
\newblock URL
  \url{http://www.sciencedirect.com/science/article/pii/S0022509620302908}.

\bibitem[Zhang et~al.(2024{\natexlab{b}})Zhang, {de Souza}, Mulvihill, Dalmas,
  Scheibert, and Xu]{zhang_nonmonotonic_2024}
Bo~Zhang, Mariana {de Souza}, Daniel~M. Mulvihill, Davy Dalmas, Julien
  Scheibert, and Yang Xu.
\newblock Non-monotonic {{Evolution}} of {{Contact Area}} in {{Soft Contacts
  During Incipient Torsional Loading}}.
\newblock \emph{Tribology Letters}, 72\penalty0 (4):\penalty0 132, November
  2024{\natexlab{b}}.
\newblock URL \url{https://doi.org/10.1007/s11249-024-01930-9}.

\bibitem[Delplanque et~al.(2022)Delplanque, Aymard, Dalmas, and
  Scheibert]{delplanque_solving_2022}
Emilie Delplanque, Antoine Aymard, Davy Dalmas, and Julien Scheibert.
\newblock Solving curing-protocol-dependent shape errors in {{PDMS}}
  replication.
\newblock \emph{Journal of Micromechanics and Microengineering}, 32\penalty0
  (4):\penalty0 045006, March 2022.
\newblock URL \url{https://doi.org/10.1088/1361-6439/ac56ea}.

\bibitem[Maugis(1992)]{maugis_adhesion_1992}
Daniel Maugis.
\newblock Adhesion of spheres: {{The JKR-DMT}} transition using a dugdale
  model.
\newblock \emph{Journal of Colloid and Interface Science}, 150\penalty0
  (1):\penalty0 243--269, April 1992.
\newblock URL
  \url{https://www.sciencedirect.com/science/article/pii/002197979290285T}.

\bibitem[Burris and Sawyer(2009)]{burris_addressing_2009}
D.~L. Burris and W.~G. Sawyer.
\newblock Addressing {{Practical Challenges}} of {{Low Friction Coefficient
  Measurements}}.
\newblock \emph{Tribology Letters}, 35\penalty0 (1):\penalty0 17--23, July
  2009.
\newblock URL \url{https://doi.org/10.1007/s11249-009-9438-2}.

\bibitem[Lu et~al.(2018)Lu, Wood, Gee, Wang, and Pfleging]{lu_novel_2018}
Ping Lu, Robert J.~K. Wood, Mark~G. Gee, Ling Wang, and Wilhelm Pfleging.
\newblock A {{Novel Surface Texture Shape}} for {{Directional Friction
  Control}}.
\newblock \emph{Tribology Letters}, 66\penalty0 (1):\penalty0 51, February
  2018.
\newblock URL \url{https://doi.org/10.1007/s11249-018-0995-0}.

\bibitem[Bigerelle et~al.(2007)Bigerelle, Mazeran, and
  Rachik]{bigerelle_first_2007}
M.~Bigerelle, P.~E. Mazeran, and M.~Rachik.
\newblock The first indenter-sample contact and the indentation size effect in
  nano-hardness measurement.
\newblock \emph{Materials Science and Engineering: C}, 27\penalty0
  (5):\penalty0 1448--1451, 2007.
\newblock URL
  \url{https://www.sciencedirect.com/science/article/pii/S092849310600333X}.

\bibitem[Guillonneau et~al.(2012)Guillonneau, Kermouche, Bec, and
  Loubet]{guillonneau_extraction_2012}
G.~Guillonneau, G.~Kermouche, S.~Bec, and J.-L. Loubet.
\newblock Extraction of {{Mechanical Properties}} with {{Second Harmonic
  Detection}} for {{Dynamic Nanoindentation Testing}}.
\newblock \emph{Experimental Mechanics}, 52\penalty0 (7):\penalty0 933--944,
  September 2012.
\newblock URL \url{https://doi.org/10.1007/s11340-011-9561-5}.

\bibitem[Gavara(2016)]{gavara_combined_2016}
N{\'u}ria Gavara.
\newblock Combined strategies for optimal detection of the contact point in
  {{AFM}} force-indentation curves obtained on thin samples and adherent cells.
\newblock \emph{Scientific Reports}, 6\penalty0 (1):\penalty0 21267, February
  2016.
\newblock URL \url{https://doi.org/10.1038/srep21267}.

\bibitem[Lucas et~al.(1998)Lucas, Oliver, and Swindeman]{lucas_dynamics_1998}
B.~N. Lucas, W.~C. Oliver, and J.~E. Swindeman.
\newblock The {{Dynamics}} of {{Frequency-Specific}}, {{Depth-Sensing
  Indentation Testing}}.
\newblock \emph{MRS Proceedings}, 522:\penalty0 3--14, 1998.

\bibitem[Oliver et~al.(2023)Oliver, Dalmas, and
  Scheibert]{oliver_adhesion_2023}
C.~Oliver, D.~Dalmas, and J.~Scheibert.
\newblock Adhesion in soft contacts is minimum beyond a critical shear
  displacement.
\newblock \emph{Journal of the Mechanics and Physics of Solids}, 181:\penalty0
  105445, 2023.
\newblock URL
  \url{https://www.sciencedirect.com/science/article/pii/S0022509623002491}.

\bibitem[Acito et~al.(2023)Acito, Dancette, Scheibert, Oliver, Adrien, Maire,
  and Dalmas]{acito_use_2023}
Vito Acito, Sylvain Dancette, Julien Scheibert, Cristobal Oliver, J{\'e}rome
  Adrien, Eric Maire, and Davy Dalmas.
\newblock On the use of in situ {{X}}-ray computed tomography for soft contact
  mechanics.
\newblock \emph{European Journal of Mechanics - A/Solids}, 101:\penalty0
  105057, September 2023.
\newblock URL
  \url{https://www.sciencedirect.com/science/article/pii/S0997753823001493}.

\bibitem[Dancette et~al.(2026)Dancette, Acito, Adrien, Lachambre, Maire,
  Scheibert, and Dalmas]{dancette_combining_2026}
Sylvain Dancette, Vito Acito, J{\'e}rome Adrien, Jo{\"e}l Lachambre, Eric
  Maire, Julien Scheibert, and Davy Dalmas.
\newblock Combining {{X}}-ray tomography and digital volume correlation to
  measure {{3D}} bulk mechanical fields in {{Hertzian}} contact mechanics.
\newblock \emph{Mechanics Research Communications}, 154:\penalty0 104671, 2026.
\newblock URL
  \url{https://www.sciencedirect.com/science/article/pii/S0093641326000613}.

\bibitem[Nagy(2024)]{nagy_determination_2024}
Norbert Nagy.
\newblock Determination of solid-liquid adhesion work on flat surfaces in a
  direct and absolute manner.
\newblock \emph{Scientific Reports}, 14\penalty0 (1):\penalty0 29991, December
  2024.
\newblock URL \url{https://doi.org/10.1038/s41598-024-81710-6}.

\end{thebibliography}

\end{document}